\documentclass[useAMS,usenatbib]{mn2e}

\include{epsf}
\usepackage{graphicx}
\usepackage{times}
\newcommand{\hi}{H\,{\sc i}}
\newcommand{\htwo}{H$_{2}$}

\newcommand{\msol}{\mbox{${\rm M}_\odot$}}
\newcommand{\hubble}{\mbox{$\rm km\, s^{-1}\, Mpc^{-1}$}}
\newcommand{\kms}{\mbox{$\rm km\, s^{-1}$}}

\newcommand{\nhi}{\mbox{$N_{\rm HI}$}}
\newcommand{\nhtwo}{\mbox{$N_{\rm H_2}$}}
\newcommand{\ohi}{\mbox{$\Omega_{\rm HI}$}}

\newcommand{\mgii}{Mg\,{\sc ii}}
\newcommand{\caii}{Ca\,{\sc ii}}
\newcommand{\lya}{Ly\,{$\alpha$}}
\newcommand{\icmsq}{\mbox{$ \rm cm^{-2}$}}

\title[HI absorption at low and intermediate redshifts]{New H{\sc i} 21-cm absorbers at low and intermediate redshifts}
\author[M.A. Zwaan et al.]{M. A. Zwaan,$^{1}$\thanks{email:mzwaan@eso.org}
J. Liske,$^{1,2}$
C. P{\'e}roux,$^{3}$
M. T. Murphy,$^{4}$
N. Bouch{\'e},$^{5}$
S. J. Curran,$^{6}$  \newauthor
and A. D. Biggs$^{1}$\\
\\
$^{1}$ European Southern Observatory, Karl-Schwarzschild-Str. 2, 85748 Garching
     b. M{\"u}nchen, Germany\\
$^{2}$ Hamburger Sternwarte, University of Hamburg, Gojenbergsweg 112, 21029 Hamburg, Germany\\
$^{3}$ Aix Marseille Universit{\'e}, CNRS, LAM (Laboratoire d'Astrophysique de Marseille) UMR 7326, 13388, Marseille, France\\
$^{4}$ Centre for Astrophysics and Supercomputing, Swinburne University of Technology, Hawthorn, Victoria 3122, Australia\\
$^{5}$ CNRS, Institut de Recherche en Astrophysique et Plan{\'e}tologie (IRAP) de Toulouse, 14 Avenue E. Belin F-31400 Toulouse, France\\
$^{6}$ School of Chemical and Physical Sciences, Victoria University of Wellington, PO Box 600, Wellington 6140, New Zealand
     }

\begin{document}

%\date{Accepted 1988 December 15. Received 1988 December 14; in original form 1988 October 11}

\pagerange{\pageref{firstpage}--\pageref{lastpage}} \pubyear{2015}

\maketitle

\label{firstpage}

\begin{abstract}
We present the results of a survey for intervening \hi\ 21-cm  absorbers at intermediate and low redshift ($0<z<1.2$). For our total sample of 24 systems, we obtained high quality data for 17 systems, the other seven being severely affected by radio frequency interference (RFI). Five of our targets are low redshift ($z<0.17$) optical galaxies with small impact parameters ($<20$~kpc) toward radio-bright background sources. Two of these were detected in 21-cm absorption, showing narrow, high optical depth absorption profiles, the narrowest having a velocity dispersion of only 1.5~\kms, which puts an upper limit on the kinetic temperature of $T_{k}<270$~K. Combining our observations with results from the literature, we measure a weak anti-correlation between impact parameter and integral optical depth in local ($z<0.5$) 21-cm absorbers. Of eleven \caii\ and \mgii\ systems searched, two were detected in 21-cm absorption,  and six were affected by RFI to a level that precludes a detection. For these two systems at $z\sim 0.6$ we measure spin temperatures of $T_s=(65\pm17)$~K and $T_s>180$~K. A subset of our systems were also searched for OH absorption, but no detections were made.

\end{abstract}


\section{Introduction}
Observing atomic hydrogen in absorption is an extremely powerful probe of galaxies at all redshifts: the detection sensitivity is not dependent on many of the properties of the system under study but  depends rather on the unrelated characteristics of the background source. Detection of the \lya\ line with absorption techniques in the optical and ultra-violet wavelengths has been very successful in the past \citep[e.g.,][and references therein]{Peroux2003a,Prochaska2005a,Meiring2011a}, providing information on the physical state of a wide variety of media: from the low density intergalactic medium to high redshift galaxies. The damped \lya\ (or DLA) systems in particular, with \hi\ column densities in excess of $\nhi=2\times 10^{20}~\icmsq$ are thought to trace gas-bearing galaxies and contain most of the neutral hydrogen atoms at all redshifts \citep[e.g.,][]{Wolfe2005a}. Indeed, these absorption line systems provide the most detailed information on the dense, pre-star-forming interstellar medium in galaxies at $z>0$. However, DLAs suffer from  selection effects associated with optically selected background sources against which the redshifted \lya\ lines are observed, such as limitations due to the atmospheric cut-off at $z=1.65$ and the potential loss of dusty DLA systems due to the selection of quasars in optical colours \citep{Vladilo2008a,Curran2011a}. 

The \hi\ hyperfine 21-cm line presents several advantages over \lya: it does not saturate and the radio continuum is unaffected by dust. Furthermore, the 21-cm line enables a determination of the kinematics and gas distribution in the intervening absorbers, which should then allow us to directly link these systems to their $z=0$ galaxy counterparts \citep[e.g.,][]{Briggs2001a}.  Comprehensive reviews of  21-cm absorption line studies of redshifted high column density gas are given by \citet{Kanekar2004b} and more recently by \citet{Morganti2015a}. 
The combination of \hi\ 21-cm and \lya\ absorption spectra yields a measurement of the \hi\ spin temperature \citep[e.g.,][]{Wolfe1979a}, which is a good indication of the kinetic temperature of cold clouds. Hence, \hi\ spin temperatures measured over a range of redshifts provide information on the evolution of the temperature of dense gas over cosmic time \citep[e.g.][]{Kanekar2003b,Kanekar2014a}. {There is evidence of a correlation between spin temperature and redshift, indicating a lower fraction of the cold neutral medium (CNM) in the ISM of galaxies at higher redshifts \citep[$z>2$;][]{Kanekar2014a}, although \citet{Curran2005a}, \citet{Curran2012a} and \citet{Curran2006a} have argued that the apparent evolution of spin temperature is largely a geometrical effect.} Despite the clear advantages of the 21-cm line, it has one marked complication:  there is a degeneracy between spin temperature and covering factor, which implies that high-resolution radio continuum data are needed to understand the absorption profile in detail.

Because of the atmospheric cut-off for \lya, very few optically selected DLAs are currently known at redshifts $z< 1$ \citep[e.g.,][]{Meiring2011a,Som2015a}. 
21-cm absorption studies are potentially a good means of increasing the sample of low redshift DLAs as these studies are relatively successful: there are currently $\sim 20$ detections of intervening \hi\ 21-cm absorption at redshifts $z<1$ \citep[][and references therein]{Curran2010b, Gupta2013a, Ellison2012a}. 
A more complete sampling of low-redshift \hi\ absorbers will eventually help in accurately determining the evolution of the cosmic mass density of neutral hydrogen \ohi\ over half the age of the universe, especially when the results of large scale 21-cm absorption line surveys from the Square Kilometre Array and its pathfinders will become available. The star formation rate density and the optical luminosity density are much better constrained over this redshift range ($z<1$), and they are observed to decline rapidly \citep[e.g.,][]{Hopkins2006a,Madau2014a}. To fully understand the evolution of the stellar and interstellar content of the universe, it is essential that \ohi\ is also  measured more precisely. Larger samples of \hi\ absorbers also assist a more solid interpretation of the evolution of cosmic metallicity at low redshifts, which is currently based on $\sim$eight $z<1$ DLAs with good metal abundance measurements \citep{Kulkarni2005a,Kulkarni2007a,Som2015a}. 

Also on much smaller physical scales low redshift 21-cm absorption line studies are important. The study of apparent pairs of foreground galaxies and background radio-bright quasars can constrain the covering factor and clumpiness of cold gas in galaxies \citep[see e.g.,][]{Reeves2015a}. In particular, very long baseline interferometry (VLBI) 21-cm absorption line studies allow measurements of small scale structure in the ISM if the background radio source is sufficiently resolved on VLBI scales \citep{Srianand2013a}. In order to find these interesting cases that show spatially resolved \hi\ absorption, searches for low-redshift \hi\ absorbers are essential. 

In this paper we present the results of a Green Bank Telescope\footnote{The National Radio Astronomy Observatory is a facility of the National Science Foundation operated under cooperative agreement by Associated Universities, Inc.} (GBT) survey for new \hi\ 21-cm absorbers at low and intermediate redshifts. We selected our targets to be either known \caii\ or \mgii\ absorbers, or galaxies with small impact parameters to  radio bright background sources. 

A complementary part of our survey consisted of searching for OH absorption lines with the GBT at low and intermediate redshifts. The systems selected to be observed in the OH 1667 MHz line are a subset of those observed in the 21-cm line, concentrating on the galaxies with small impact parameters to background radio sources. The purpose of this survey was to identify molecular absorption lines in the centres of galaxies, where it is known that the ratio of atomic to molecular gas density is higher \citep[e.g.,][]{Leroy2009b}. Previous observations of the OH line in known molecular absorbers indicate that OH is an excellent tracer of molecular gas in intermediate redshift absorbing systems, and its detection also enables an accurate \htwo\ column density measurement \citep{Liszt1999a,Kanekar2005a}.

The approach of this paper is to first concentrate on the \hi\ part of the survey. Section 2 describes in detail the selection of our sample and in section 3 we report on the data taking and data reduction. In section 4 we describe the detected \hi\ 21-cm absorption systems individually in some detail and in section 5 we discuss the statistics of low-redshift \hi\ absorbers. The results and implications of the OH absorption part of our survey is described in section 6. Finally, section 7 presents the conclusions.  Throughout this paper we adopt $H_0=75 \,\hubble$, $\Omega_\Lambda=0.70$, and $\Omega_M=0.30$.

\section{HI 21cm Target selection}
Three different selection techniques were used to identify candidates for 21-cm absorption observations. We briefly describe all three methods here. The complete list of targets is presented in Table \ref{obstabel.tab}, where the column 'selection' indicates which of the selection methods was used to include the target. Note that some targets meet the selection criteria of more than one method.

\begin{table*}
\label{obstabel.tab}
\caption{Properties of the targets selected for HI and OH observations.
Column 1: the name of the background radio source. Systems for which \hi\ lines are detected are indicated in bold face; column 2: J2000 coordinates of the background radio source; column 3: emission redshift of the background source, if available; column 4: expected redshift of the absorber. In the case of \mgii, \caii\ and \lya\ absorbers, $z_{\rm abs}$ is the measured absorber redshift and in the case of known galaxies the redshift of the galaxy is given. $^*$ denotes an associated system; column 5: measured 1.4~Jy flux density as given in FIRST; column 6: selection method with impact parameter in kpc given if appropriate; column 7: \mgii\ $W_0^{2796}$ equivalent width in \AA\ measured from SDSS data; column 8: similar for \caii\ $W_0^{3934}$; column 9: reference 
(1) \citet{Chen2001a},
(2) \citet{Churchill2001a},
(3) \citet{Rao2006a},
(4) \citet{York2006a},
(5) \citet{Sardane2014a},
(6) \citet{Prochter2006a}.
\newline 
For the \mgii\ $W_0^{2796}$ measurements the most recent values are taken from either \citet{Rao2006a}, \citet{Rao2011a} or \citet{Quider2011a}. Note that 'small $b$' refers to  absorption line systems identified with small impact parameter galaxies, whereas 'SDSS galaxy' refers to low redshift galaxies without prior detection of associated absorption lines.
}
\begin{tabular}{l l l l l l l l l}
\hline
Name &Coordinates (J2000) &$z_{\rm em}$ &$z_{\rm abs}$ or $z_{\rm gal}$ &S$_{\rm 1.4}$&Selection & \mgii\ $W_0^{2796}$ & \caii\ $W_0^{3934}$ & Reference\\
& \hspace{1mm}h \hspace{1mm}m\hspace{2mm}s  \hspace{7mm}$^\circ$ \hspace{2mm}$'$ \hspace{2mm}$''$ & & & (Jy) & (impact par [kpc]) &  (\AA) & (\AA) &  \\
\hline\hline
FBQS J0154-0007		&01  54  54.36  $-$00 07  23.3 	&1.826940 &1.180    	&0.26	&\mgii\			 			& 1.399 && 	 6 \\
SBS 0846+513			&08  49  57.98  $+$51 08  29.0 	&0.583715 &0.0734   	&0.34	&SDSS galaxy (19.0)	 	&&&this work		 \\
{\bf GB6 J0855+5751}	&08  55  21.36  $+$57 51  44.1 	&. . .	  &0.0260   		&0.62	&SDSS galaxy (9.1)	 		&&&this work		 \\
{\bf B3 0927+469}		&09  30  35.08  $+$46 44  08.7 	&2.032000 &0.6213   	&0.31	&\caii\ /\mgii\	 	 	& 3.255 &0.600	& 5\\
4C +04.33				&09  46  42.42  $+$04 19  00.3 	& . . .	  &0.0877       	&0.55	&SDSS galaxy (20.0)	 	&&&this work		 \\
GB6 J1103+1114		&11  03  34.79  $+$11 14  43.0 	&1.72934  &0.7176   	&0.27	&\caii\	     		 		& 3.470 &0.307& 5\\
Q1148+387			&11  51  29.37  $+$38 25  52.4 	&1.302510 &0.5530   	&0.50	&small $b$/\mgii\ (18.9)	& 0.482 && 2		 \\
B1213+590			&12  16  04.72  $+$58 43  33.3  	&1.453360 &0.7246   	&0.38	&\caii\ /\mgii\	 	 	& 2.786 &0.430	& 5\\
B1239+606			&12  41  29.59  $+$60 20  41.3  	&2.068660 &1.2381   	&0.43	&\caii\	 		 			&&1.034& 5\\
PG1241+176			&12  44  10.83  $+$17 21  04.5  	&1.273000 &0.5507   	&0.43	&small $b$/\mgii\ (20.0)	& 0.570 && 2		 \\
SBS 1307+562			&13  09  09.75  $+$55 57  38.2  	&1.629370 &0.5096   	&0.29	&\mgii\	      		 		& 2.355  && 6\\
B3 1325+436			&13  27  20.98  $+$43 26  28.0 	&2.084390 &0.9539   	&0.66	&\caii\	 		 			& 2.020 &0.901	& 5\\
{\bf 4C +57.23} 			&13  54  00.12  $+$56 50  04.7 	&. . .	  &0.0955   	  	&0.61	&SDSS galaxy (11.7)	 		&&&this work		 \\
{\bf J1431+3952}		&14  31  20.54  $+$39 52  41.5  	&1.215400 &0.6022   	&0.21	&\mgii\	 		 	& 2.474 && 6\\
SDSS 1445+0347		&14  45  53.47  $+$03 47  32.5  	&1.300080 &1.248    	&0.28	&\mgii\	 		 			& 1.01 && 6	 \\
PKS 1545+21			&15  47  43.54  $+$20 52  16.6	&0.264300 &0.2657$^*$ &2.4	&small $b$/\lya\ (11.5)	 		&&& 1			 \\
PKS 1602-00			&16  04  56.14  $-$00 19  06.9  	&1.628760 &1.325    	&1.0	&\mgii\	     		 		& 0.67 && 4	 \\
3C336				&16  24  39.09  $+$23 45  12.2  	&0.927398 &0.9310   	&2.5	&small $b$/\lya\ (19.0)	 	&&& 1			 \\
3C336-1				&16  24  39.09  $+$23 45  12.2  	&0.927398 &0.8920   	&2.5	&small $b$/\mgii\ (20.0)	& 1.622 && 2		 \\
PKS 2003-025			&20  06  08.49  $-$02 23  35.1  	&1.457000 &1.2116   	&2.1	& \mgii\ 	 	 	& 2.65 && 3	 \\
PKS 2135-14			&21  37  45.17  $-$14 32  55.8 	&0.200300 &0.1996   	&3.6	&small $b$/\lya\ (16.5)	 	&&& 1 		 \\
PKS 2149+212	 		&21  51  45.95  $+$21 30  13.8  	&1.538500 &0.9114   	&0.9	& $\log \nhi=20.7$	 	 		&&& 3	 \\
PKS 2330+005			&23  33  13.17  $+$00 49  11.9  	&0.169989 &0.1699   	&0.35	&SDSS galaxy (17.1)	 	&&&this work		 \\
PKS 2355-106			&23  58  10.88  $-$10 20  08.6 	&1.639110 &1.172    	&0.77	&\mgii\        		 	& 1.569 && 6	 \\
\hline
\end{tabular}
\end{table*}%

Our first set of objects consists of \mgii\ absorbers. 
It has been known for many years that a large fraction of \mgii\ absorbers show 21-cm absorption \citep[e.g.][]{Briggs1983a,Lane2000a}, and it has been argued that this fraction rises when those \mgii\ absorbers are selected that show the highest equivalent widths. In particular, \citet{Lane2000a} claimed that almost all \mgii\ systems with $W_0(2796)>2$~\AA\ show 21-cm absorption and
\citet{Curran2010a} find a weak correlation between the 21-cm line strength normalised by the column density and the \mgii\ equivalent width. However, \citet{Kanekar2009a} do not confirm a statistically significant relation between \mgii\ equivalent width and 21-cm detectability and report an overall detection rate of 21-cm absorption in 25 per cent of the \mgii\ systems. 
 The lack of a strong correlation between \mgii\ absorption properties and 21-cm absorption is somewhat surprising, as it has been shown that \hi\ column density measured through \lya\ absorption does correlate with \mgii\ equivalent width. \cite{Rao2006a} showed that the fraction of DLA systems among \mgii\ systems rises to 42 per cent when those systems are selected for which $W_0(2796)>0.5$~\AA\ and $W_0(2796)/W_0($Fe$ 2600)>2$ and \cite{Menard2009a}  report an $8\sigma$ correlation between the mean \lya\ \hi\ column density and $W_0(2796)$ over the redshift range $0.5<z<1.4$.
 
In order to increase the sample of \mgii\ absorbers with \hi\ 21-cm absorption, we  cross-correlated \mgii\ absorbers identified in the Sloan Digital Sky Survey (SDSS) with  background sources  brighter than 200 mJy at 1.4 GHz in the FIRST data base \citep{Becker1995a}. The \mgii\ sample was drawn from the compilations of \citet{Rao2006a}, \citet{York2006a},  \citet{Prochter2006a}, and \citet{Sardane2014a}. Four systems from this list fulfil the DLA criterion of \cite{Rao2006a} and have a background source brighter than 200 mJy at 1.4 GHz. We added to this sample one absorber from the list of \citet{Rao2006a}, which is a strong \lya\ absorber. The total sample of \mgii\ absorbers is listed in Table~\ref{obstabel.tab}, with the original reference indicated. 

\caii\ absorbers form our second group of objects.   These systems probably form a  dustier and more chemically evolved subset of DLAs. \citet{Wild2005a} argue that the average $E(B-V)$ values and Zn{\sc II} column densities combined with conservative assumptions regarding metallicities and dust-to-gas ratios point to \caii\ systems having \hi\ column densities above the DLA limit.  Selection by \caii\ absorption is an effective way of identifying high column densities of neutral hydrogen, and thus samples of DLAs at low redshift. 
The SDSS provides an ideal database of quasar spectra to search for these systems. Here, we use the method described by \citet{Zych2009a}  to select \caii\ absorbers.  Five of these systems are found to be in sight lines towards radio loud background sources ($>200$ mJy). 

The third group of targets consists of galaxies with small impact parameters to background radio sources. For this list we assembled targets using two methods: {\em i)} we assembled galaxies from the literature that are identified with known quasar absorption lines at intermediate redshift and have a radio loud background source; {\em ii)} we cross-correlated SDSS galaxies with spectroscopic redshifts with the FIRST catalogue to identify quasar galaxy pairs. The impact parameters of the selected systems range from 9 to 20~kpc.
Table \ref{obstabel.tab} summarizes the details of our sample. 

\section{Observations and data reduction}
Observations were carried out with the NRAO Green Bank Telescope (GBT) during various observing runs between April 2006 and September 2007. For the back-end we used the spectral processor, with a bandwidth of 1.25 or 2.5 MHz, two polarizations, 1024 channels and a 5s integration time. This setup results in a spectral resolution of approximately $0.5(1+z) \rm km\, s^{-1}$ or $1.0(1+z) \rm km\, s^{-1}$. The receivers used for this program were the prime focus instruments PF1-3 (600 MHz), PF1-4 (800 MHz), and PF2 (1 GHz), with system temperatures of $T_{\rm sys}=48$K, $25$K, and $22$K, respectively, and the L-band receiver with $T_{\rm sys}=20$K. The observations were taken in standard on-off mode, with 5 min on and 5 min off scans. On-source integration times ranged from 1 to 5 hours, depending on the flux of the background radio continuum source. 

Unfortunately, observations in the frequency range much below 1400 MHz are plagued by intermittent strong radio frequency interference (RFI). The targets that could not be observed because of strong RFI were usually visited again several days later. For four targets we have not been able to acquire any good quality data, namely  B1239+606, SDSS 1445+0347, PKS 2003-025 and PKS 2355-106, at sky frequencies of 634.6, 631.8, 642.2, and 653.8~MHz, respectively. All these cases of strong RFI occurred with observations using the PF1-3 receiver, illustrating the difficulty of achieving high quality data around 600 MHz (corresponding to the \hi\ 21-cm line at $z\sim 1.2$) with the GBT.

Data reduction was carried out with the DISH data reduction tool, which is based on the aips++ package. DISH allows for a flexible editing of the data in order to remove data sections with RFI or obvious correlator problems. Initial data editing was done with the autoflag tool, after which all data were scrutinised by eye. Data taken in one session were median filtered and a baseline was removed by making a third-order polynomial fit to the line-  and RFI-free channels. Data taken over different days were averaged and any residual baselines were removed by means of a final second-order polynomial fit.

\section{Detected \hi\ lines}
The strength of an \hi\ absorption feature per observed frequency channel can be expressed as the optical depth, defined as 
\begin{equation} \label{tau.eq}
\tau=-\ln (1-\frac{\Delta S}{fS}),
\end{equation}
where $S$ is the background source flux at the observed frequency, $\Delta S$ is the absorbed flux, and
$f$ is the 'covering factor', i.e., the flux-weighted fraction of the continuum source covered by the absorber. In the optically thin case, where $\Delta S/S<0.3$, equation~\ref{tau.eq} reduces to 
\begin{equation} 
\tau\approx \frac{\Delta S}{fS}.
\end{equation}

 By replacing $\Delta S$ with the rms flux density of the spectrum, we can calculate the rms of $\tau$, or $\sigma_\tau$.
We conservatively assume that an absorption line is significant if it is detected with a strength of $5\sigma$ over a velocity width of 10 $\rm km\, s^{-1}$. The integrated optical depth corresponding to this detection limit is defined as $\tau dV_{5,10}$.

The \hi\ column density \nhi\ can be calculated from the observed optical depth using 
\begin{equation}\label{eq2.eq}
\nhi=1.823\times 10^{18} (T_s/f) \int \tau dV\, [\rm cm^{-2}],
\end{equation}
where $T_s$ is the harmonic mean spin temperature of the absorbing gas and $f$ is the covering factor. VLBI continuum observations can be used to estimate this covering factor, by calculating the fraction of the total measured flux density contributed by the core region. The assumption is then that the \hi\ absorption arises against the core\footnote{Note that in reality we have no information on the actual extent of the absorbing material, and whether it fully covers the background emission region.}. 

\begin{figure*}
\begin{center}
\includegraphics[width=16cm,trim=1.3cm 10cm 1.3cm 9.7cm]{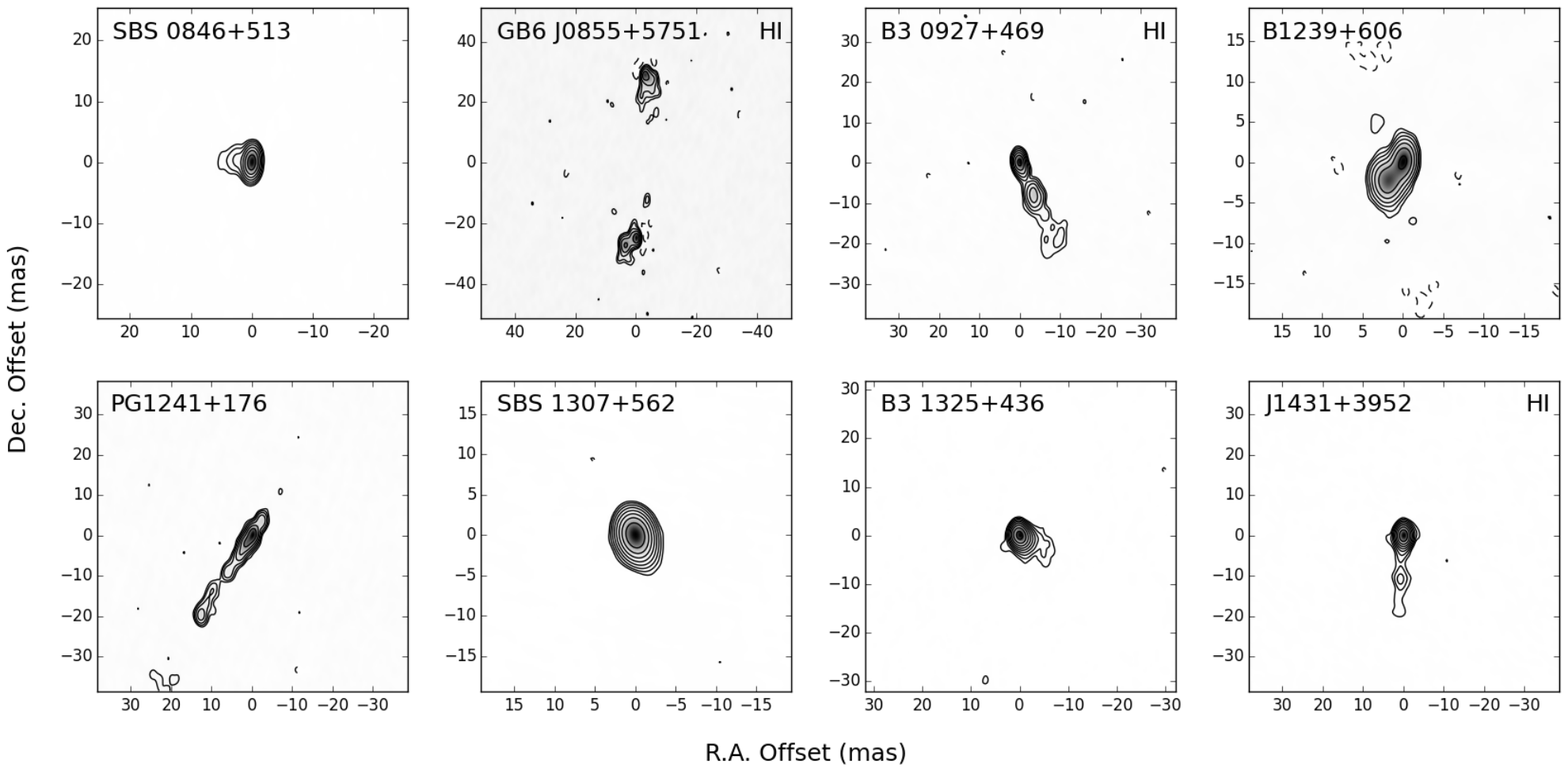}
\caption{VLBA maps of eight of the background sources from our survey. The label 'HI' in the top right corner of three maps indicates that 21-cm \hi\ absorption has been detected against these targets. The data have been taken from \citet{Helmboldt2007b}. \label{vlba.fig}}
\end{center}
\end{figure*}

Eight of our targets have been observed by \citet{Helmboldt2007b} as part of their 5~GHz VLBA Imaging and Polarimetry Survey (VIPS) of flat radio spectrum sources. Figure~\ref{vlba.fig} shows a reproduction of their VLBA maps of these eight targets. The covering factors $f$ were estimated from these maps by dividing the flux from the central point source component by the total flux measured in the map. Since these maps are made at 5~GHz, while our observations are at much lower frequencies, we note that the calculation of the covering factors is intrinsically uncertain and should be taken as an estimate only. The results are tabulated in Table~\ref{covering.tab}.

\begin{table}
\label{covering.tab}
\caption{Covering factors $f$ estimated from 5~GHz VLBA maps from \citet{Helmboldt2007b} presented in Figure~\ref{vlba.fig}.}
\centering
\begin{tabular}{l c}
\hline
name & $f$\\
\hline\hline
SBS 0846+513		& 0.97 \\
GB6 J0855+5751	& $\sim0.50$ \\
B3 0927+469		& 0.55	 \\
B1239+606		& 0.66 \\
PG1241+176		& 0.40 \\
SBS 1307+562		&  1.00 \\	
B3 1325+436		& 0.99 \\	
J1431+3952		& 0.95 \\		
\hline
\end{tabular}
\end{table}%

Neutral hydrogen 21-cm absorption features were searched for by eye in the final spectra. We identify four significant 21-cm absorbers. In the following we describe these detections is some detail. 
Table \ref{nondetections.tab} summarizes the characteristics of the reduced \hi\ spectra that do not show any \hi\ absorption. This table lists the values of $\sigma_\tau$, as well as the upper limits to the integrated optical depth $\tau dV_{5,10}$ and the channel width $dV$ in $\rm km\, s^{-1}$. A summary of the detections is given in Table~\ref{detections.tab}.

\begin{table*}
\label{nondetections.tab}
\centering
\begin{minipage}{130mm}
\caption{Parameters of the spectra in which no 21-cm absorption was detected.}
\begin{tabular}{l l l l l}
\hline
name & $\sigma_\tau$ & $\tau dV_{5,10}$ & $dV$ & comments\\
 & & (\kms) & (\kms) \\
\hline\hline
FBQS J0154-0007	&		$<0.071$	& ...				& 0.562 	& strong RFI near expected line\\
SBS 0846+513		&		$0.0046$	& $<0.17$		& 4.5		& baseline ripple, some RFI\\
4C +04.33			&		$0.0099$	& $<0.14$		& 0.449 	& \\
GB6 J1103+1114	&		$0.0069$	& $<0.16$		& 1.75	& \\
Q1148+387		&		$0.0069$	& $<0.14$		& 0.80	& \\
B1213+590		&		$0.0049$	& $<0.089$		& 0.89	&	\\
B1239+606		&		...			& ...				& 0.57	& lost in RFI\\
PG1241+176		&		$0.0080$	& $<0.17$		& 0.80	& \\
SBS 1307+562		&		$0.030$		& ...				& 0.78	& RFI exactly at expected line\\
B3 1325+436		&		$0.010$		& $<0.21$		& 1.00	& \\
SDSS 1445+0347	&		...			& ...				& 0.58	&  lost in RFI \\
PKS 1545+21		&		$0.0028$	& $<0.064$		& 0.78	&\\
PKS 1602-00		&		$0.0071$	& $<0.20$		& 2.35	& RFI near expected line \\
3C336			&		$0.0049$	& $<0.11$		& 0.95	& \\
3C336-1			&		$0.0030$	& $<0.068$		& 1.0		& \\
PKS 2003-025		&		...			& ...				& 0.57	& lost in RFI\\
PKS 2135-14		&		$0.0022$	& $<0.052$		& 0.63	&\\
PKS 2149+212	 	&		$0.0021$	& $<0.055$		& 0.99	& \\
PKS 2330+005		&		$0.012$		& $<0.20$		& 0.75	&\\
PKS 2355-106		&		...			& ...				& 0.56	& lost in RFI  but detected by \citet{Gupta2009a}\\
\hline
\end{tabular}
Note that $\tau dV_{5,10} $ indicates an upper limit to the integral optical depth, assuming a significance of $5\sigma$ over a velocity width of 10 $\rm km\, s^{-1}$.
\end{minipage}
\end{table*}%

\begin{table*}
\label{detections.tab}
\centering
\begin{minipage}{130mm}
\caption{Properties of the detected 21-cm absorbers.}
\begin{tabular}{l l l l l l}
\hline
name & $\sigma_\tau$ & $\tau_{\rm max}$ & $dV$ & $\tau dV $ & redshift\\
 & & & (\kms) & (\kms) \\
\hline\hline
GB6 J0855+5751		&		$0.0043$		& 0.24	& 0.21	& 1.02	& 0.02581  \\
B3 0927+469			&		$\sim 0.0057$	& 0.032	& 0.82	& 0.36	& 0.62146  \\
4C +57.23 			&		$0.0060$		& 0.14	& 0.45	& 3.03	& 0.09525    \\
J1431+3952			&		$0.0235$		& 0.22	& 0.83	& 4.29	& 0.60185   \\
\hline
\end{tabular}
\end{minipage}
\end{table*}%

\subsection{Intermediate redshift \hi\ absorbers}
\begin{figure}
\begin{center}
\includegraphics[width=8.0cm,trim=1.0cm 0.5cm 0.5cm 0.5cm]{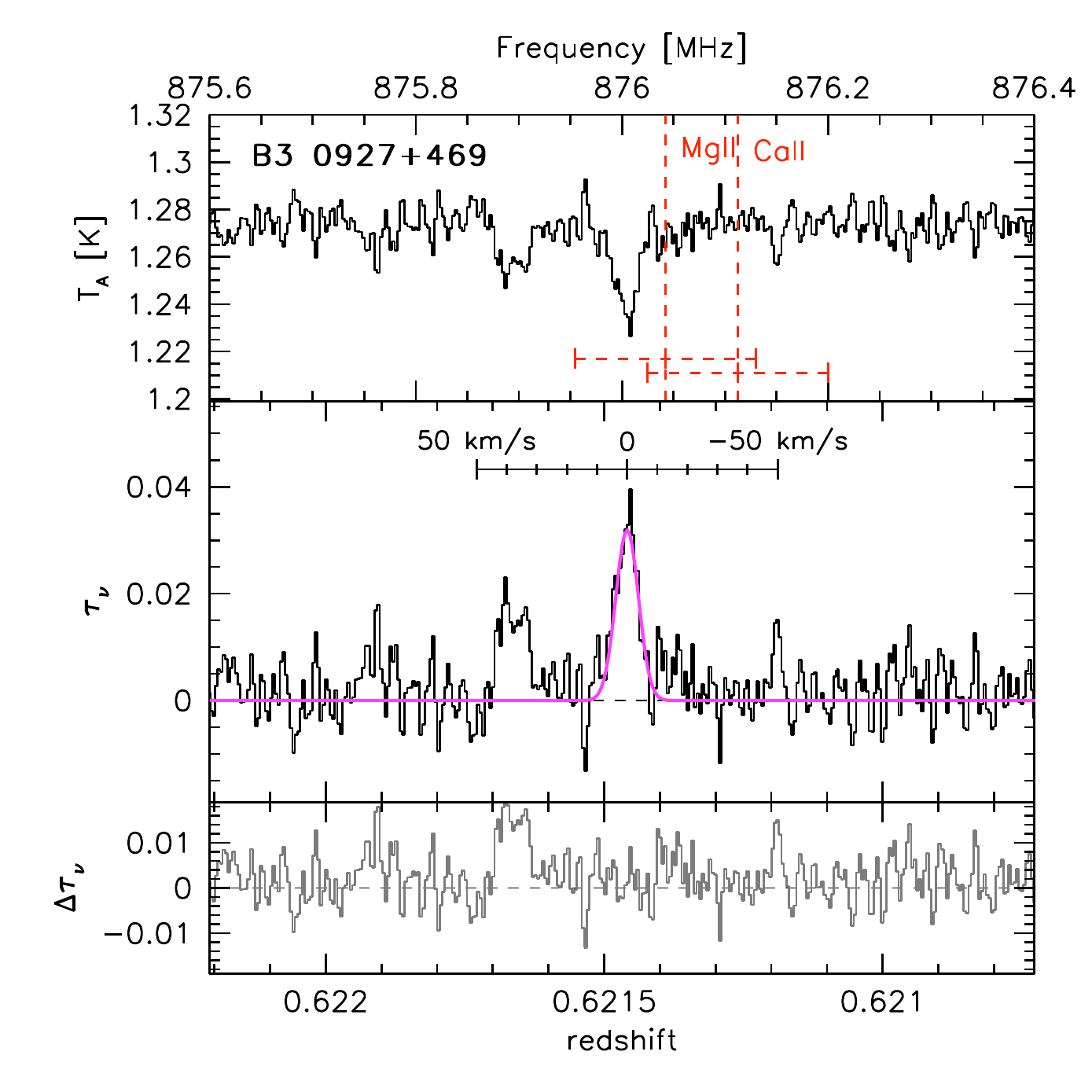}
\caption{{\em Top:\/} GBT spectrum of the \hi\ 21-cm absorber at $z=0.621$ against B3 0927+469.  RFI is severe in this spectrum, as indicated by the noisy signatures red-ward of the absorption line. Due to this strong RFI we cannot confirm whether the tentative second absorption component at a frequency of 875.9 MHz is real. The vertical lines indicate the frequencies corresponding to the \hi\ line redshifted to the \mgii\ and \caii\ lines detected in this system. The error bars give an indication of the redshift uncertainties of these lines; {\em Middle:\/} Optical depth distribution of this spectrum. The peak optical depth is  3.2 per cent. A single Gaussian fit to the absorption profile is shown as a solid line and the baseline is indicated by a dashed line. The scale bar indicates the velocity relative to the redshift corresponding to peak optical depth; {\em Bottom:\/}  The residual optical depth spectrum after subtracting the Gaussian fit. \label{0927_HI.fig}}
\end{center}
\end{figure}

\begin{figure}
\begin{center}
\includegraphics[width=8.0cm,trim=1.0cm 0.5cm 0.5cm 0.5cm]{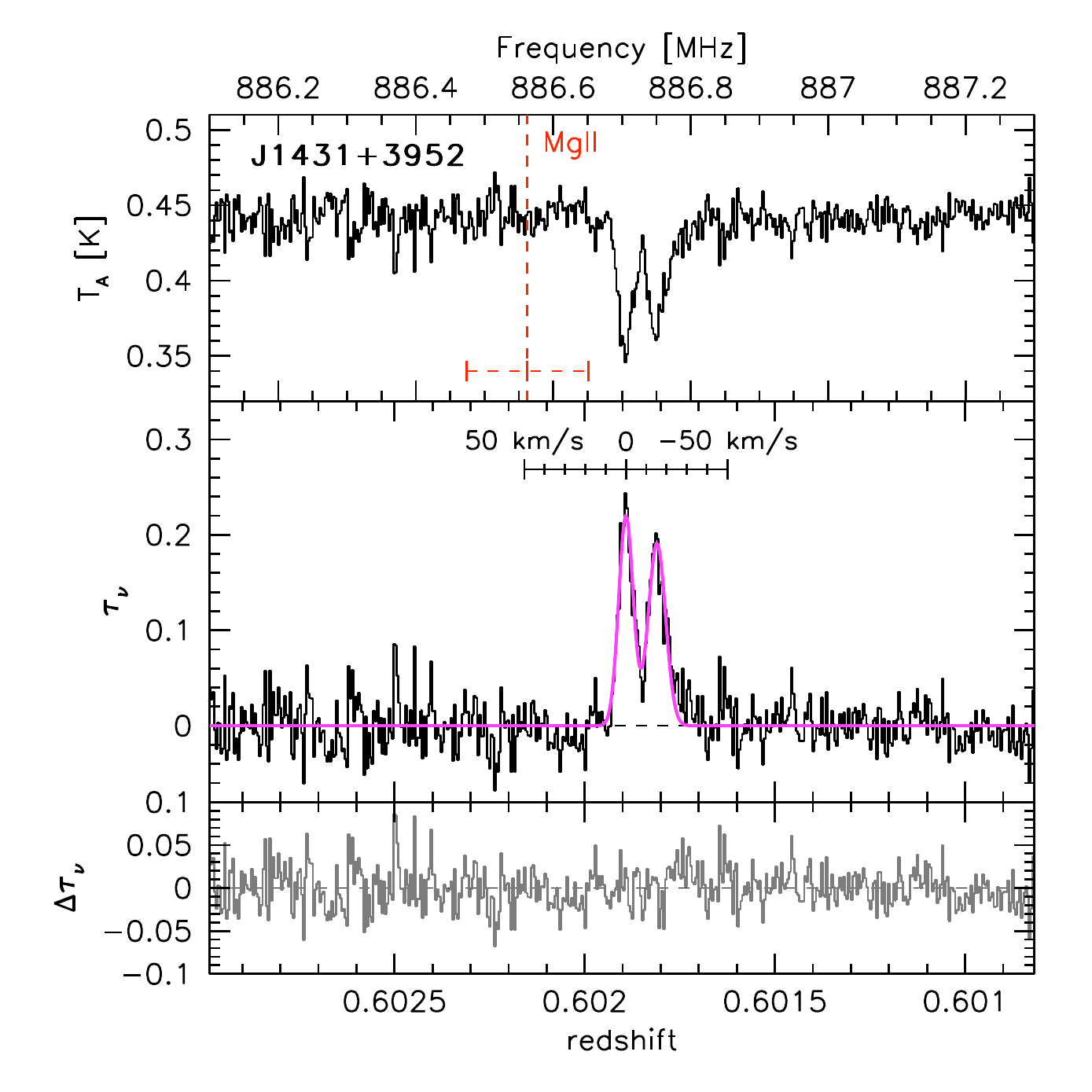}
\caption{{\em Top:\/} GBT spectrum of the \hi\ 21-cm absorber at $z=0.602$ against J1431+3952.  The vertical line indicates the frequency corresponding to the \hi\ line redshifted to the \mgii\ line detected in this system. The error bar gives an indication of the redshift uncertainties of this line; {\em Middle:\/} Optical depth distribution of this spectrum. The peak optical depth is 22 per cent. A double Gaussian fit to the absorption profile is shown as a solid line  and the baseline is indicated by a dashed line; {\em Bottom:} The residual spectrum after subtracting the double Gaussian fit.\label{1431_HI.fig}}
\end{center}
\end{figure}

\begin{figure}
\begin{center}
\includegraphics[width=3.5cm,trim=1.7cm 11.0cm 4.2cm 0.cm]{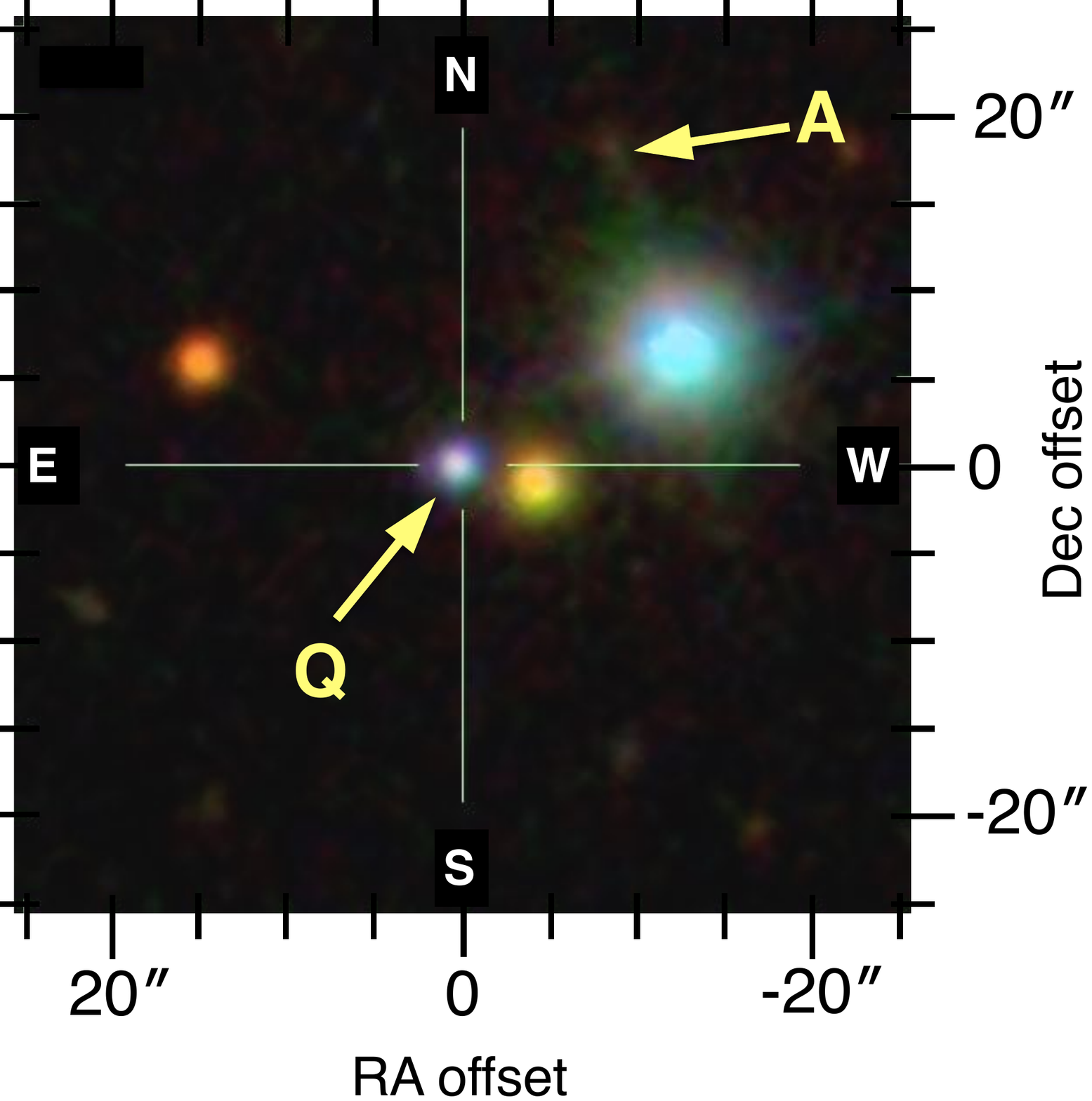}
\includegraphics[width=3.5cm,trim=0.9cm 11.0cm 5.0cm 0.cm]{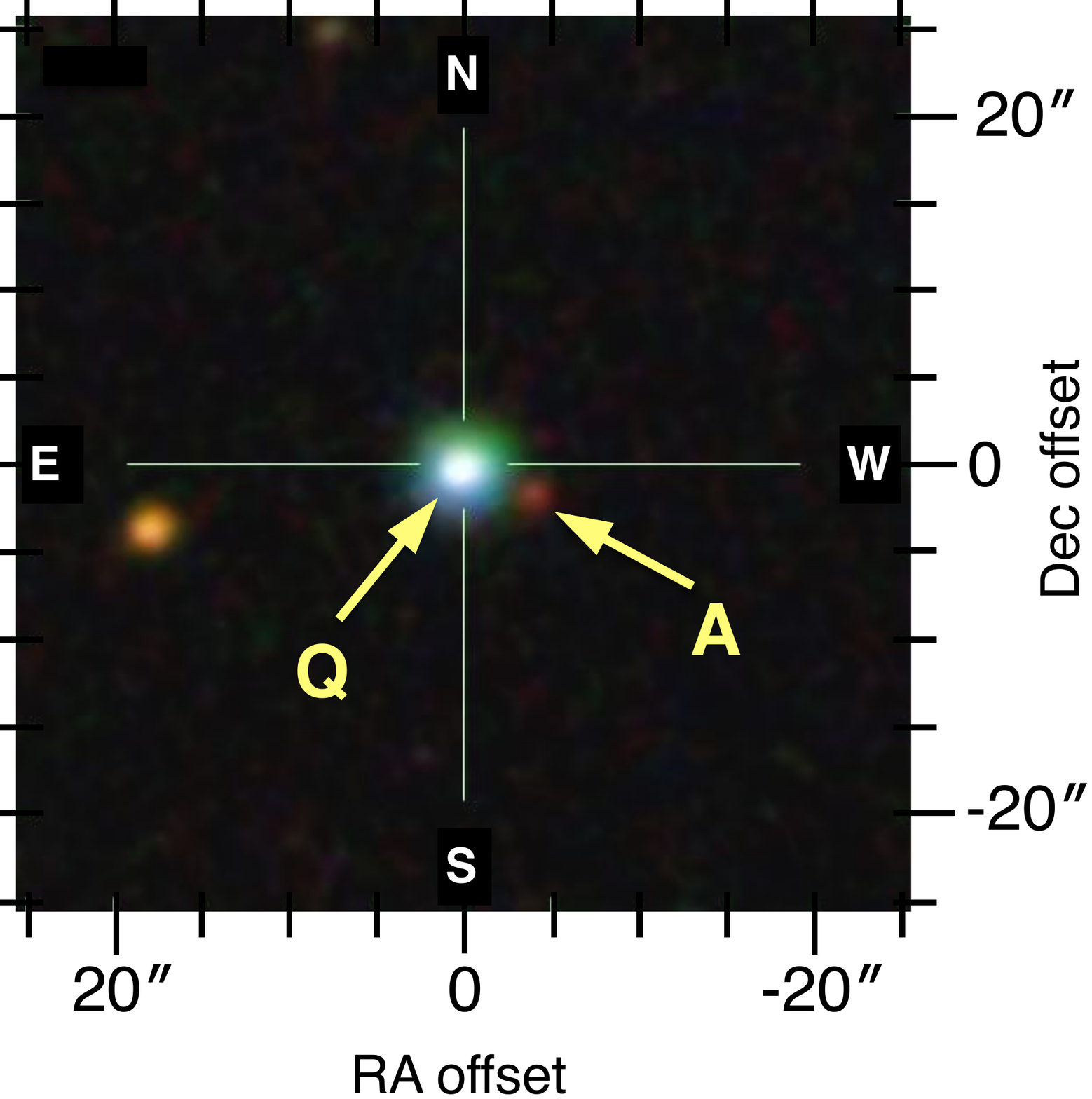}
\caption{Multicolour SDSS images of the fields around B3 0927+469 (left) and J1431+3952 (right). In both images the background source is in the centre and marked with 'Q'. The candidate absorbing galaxies that are discussed in the text are marked with 'A'. The axes are in arcsec, offset from the positions given in Table~1. The colour images are based on a combination of $g-$, $r-$, and $i-$band images, following \citet{Lupton2004a}. \label{sdssHI.fig} }
\end{center}
\end{figure}

We identify two  significant absorption features at intermediate redshifts: the \mgii-selected absorber at $z=0.602$ towards J1431+3952 and the \caii\ absorber at $z=0.621$ towards B3 0927+469. The spectra of both detections are presented in Figures \ref{0927_HI.fig} and \ref{1431_HI.fig}, where we show the antenna temperature $T_{\rm A}$ in K in the top panel, and the observed optical depth in the lower panel. In Figure \ref{sdssHI.fig} we show multicolour SDSS images of the two fields, centred on the position of the background radio sources. In the next two subsections the characteristics of these absorbers at $z\sim0.6$ are described in more detail.

\begin{figure}
\begin{center}
\includegraphics[width=7.0cm,trim=0.5cm 0.5cm 0.5cm 0.5cm]{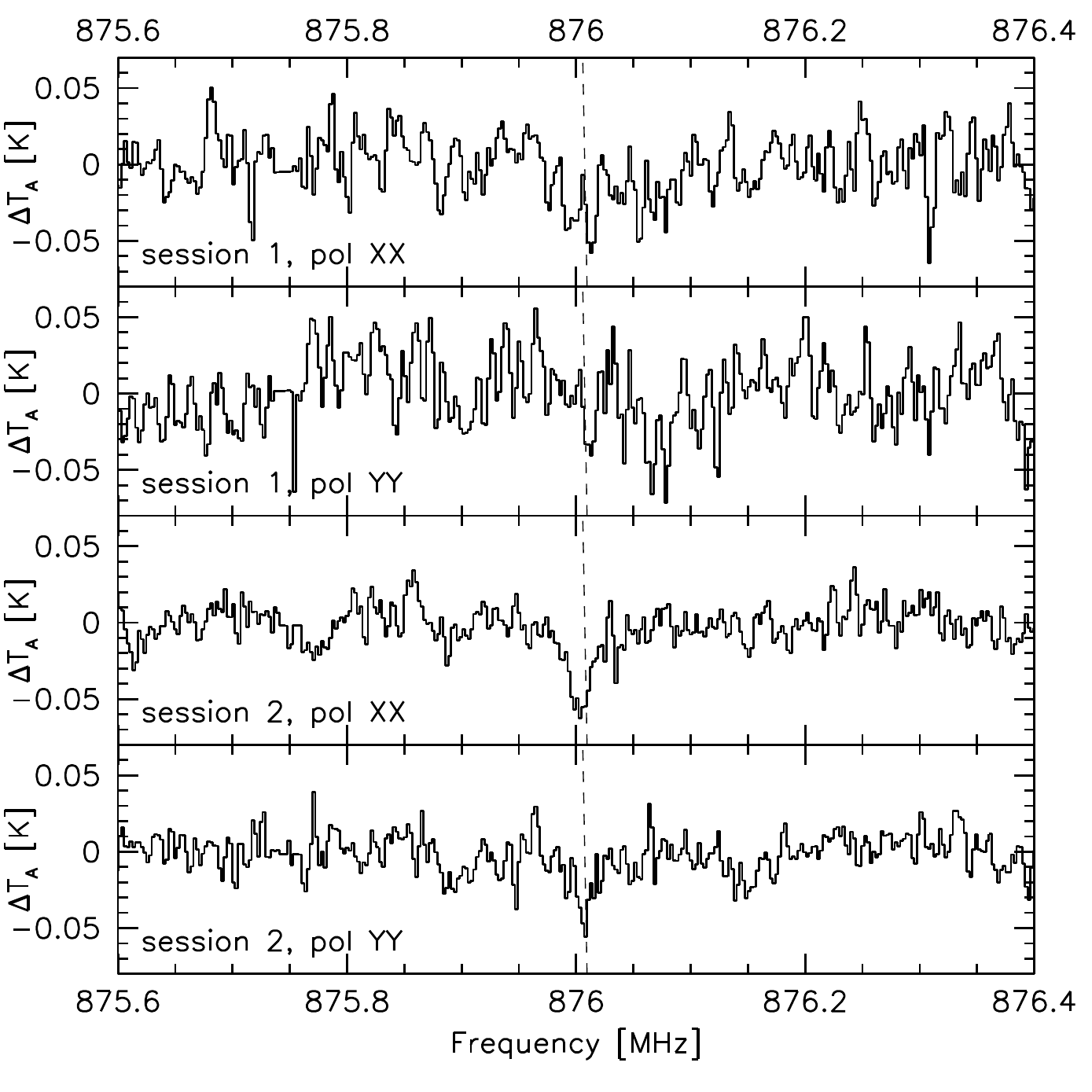}
\caption{GBT spectra of B3 0927+469 for each polarisation product (XX and YY), and for each of the observing sessions, separated by three days.  The absorption feature around 876~MHz is apparent in each of the four independent spectra. \label{0927-check.fig}}
\end{center}
\end{figure}

\subsubsection{B3 0927+469}
The absorption feature seen in the spectrum of  B3~0927+469 has a single component with a peak optical depth of 0.032. We can fit the absorption line with a single Gaussian with a centre frequency of 876.005 MHz, corresponding to a barycentric redshift of $z=0.6214$, and a velocity dispersion of $\sigma=3.8 \pm 0.8~\kms$. The integrated optical depth of the line is $\tau dV=(0.36 \pm 0.05) ~\kms$. There is tentative evidence for a second absorption component at a frequency of 875.9 MHz, but due to strong RFI we cannot confirm this detection. Note that this system was also observed with the WSRT by \citet{Gupta2012a}, but not detected, although the upper limit on the integral optical depth they quote ($\tau dV = 0.24$~\kms) is lower than the value we measure. The rms noise in our spectrum is approximately 20 per cent lower than what \citet{Gupta2012a} achieved. To confirm the reality of our detection, we plot in Figure~\ref{0927-check.fig} separate spectra for each polarisation and for each of the two observations of this target, separated by three days. It can be seen that the absorption feature is detected in each of the four independent spectra.
  
This system was first identified as a \caii\ and \mgii\ absorber in the SDSS data base. The measured redshifts of these lines are $z=0.62126$ and $z=0.62139$, respectively. The uncertainties on these measurements are approximately 30 \kms, indicating that there is no evidence for a redshift offset between the 21-cm line and the \caii\ and \mgii\ lines. 
From the 5~GHz VLBA maps of \citet{Helmboldt2007b} we estimate that $f=0.55$ (see Table~\ref{covering.tab}). The \hi\ column density of the system is therefore $1.1\times 10^{18} T_s \,\rm cm^{-2}$, which is on the low end of the distribution of known redshifted 21-cm absorbers \citep{Curran2006a}.  
Assuming that \caii\ systems are DLAs, which implies that $\nhi>2\times 10^{20}~\icmsq$, we can put a lower limit to the spin temperature of $T_s>180$~K.
The assumption that this absorber has an \hi\ column density above the DLA limit is corroborated by the results of \citet{Menard2009a}, who find a strong correlation between the \mgii\ equivalent width and the \hi\ column density. Using their relation and the measurement of \mgii\ $W_0^{2796}=3.255$~\AA\ (Table 1), we find that $\nhi=(2.4\pm 0.5)\times 10^{20}~\icmsq$. Furthermore,  \citet{Ellison2006a} and \citet{Ellison2009a} defined the so-called '$D$-index' as the ratio of the \mgii\ equivalent width and the total velocity spread of the \mgii $\lambda 2796$ profile $\times 1000$, to identify DLAs. For B3~0927+469 we find $D=8.1$, which puts it in the zone where the probability of the system being a DLA is 54 per cent \citep{Ellison2009a}.

The SDSS image of the field around B3~0927+469 in Figure~\ref{sdssHI.fig} shows several possible galaxy candidates that could be responsible for  the 21-cm absorption. We used the SDSS photometry and the Le Phare\footnote{http://lephare.lam.fr} software \citep{Arnouts1999a,Ilbert2006a} to determine photometric redshifts, and found one candidate for which the derived redshift is consistent with that of the 21-cm absorption: SDSS J093034.22+464426.8 at  $z=0.67_{-0.28}^{+0.20}$. However, the impact parameter would be $\sim 150$~kpc, which renders it very unlikely that this galaxy is hosting the gas that is causing the \hi\ absorption.

\subsubsection{J1431+3952}
After our observations of this system, the same \hi\ absorption line was detected in new GBT observations reported by \citet{Ellison2012a}. The quality of our data and the \citet{Ellison2012a} data is comparable: they report an rms noise of 4.5 mJy per 0.83~\kms\ channel, while we measure a noise level of 5.0 mJy. We measure a peak optical depth of $\tau=0.22$ in the spectrum of J1431+3952. This system shows a double structure, which we fit with parameters 
$z_1=0.60189$,
$z_2=0.60181$,
$\sigma_1=(3.5 \pm 0.8) \,\kms$,
$\sigma_2=(4.2 \pm 0.8) \,\kms$,
and peak optical depth values of 
$\tau_1=0.22$,
$\tau_2=0.19$.
The peak-to-peak velocity difference between the two peaks is $15.3 \,\kms$. For the integrated optical depth of the absorption feature we measure $(4.0 \pm 0.3) \,\kms$, while \citet{Ellison2012a} report $(3.07 \pm 0.34) \,\kms$. Also for this system the redshift of the 21-cm line is consistent with that of $z=0.60215$ for the \mgii\ system, given the uncertainty in the redshift of the latter.

This absorption line system was first identified as a \mgii\ system by \citet{Quider2011a}. \citet{Ellison2012a} use archival HST STIS measurements of the \lya\ line to measure a column density of $\log (\nhi/\icmsq)=21.2\pm 0.1$. From the 5~GHz VLBA maps of \citet{Helmboldt2007b} we estimate that the covering factor is $f=0.95$, but the 1.4~GHz maps presented in \citet{Ellison2012a}  suggest a much lower covering factor of $f\approx 0.30$.
\citet{Ellison2012a} claim that the resulting spin temperature of $T_s=(90 \pm 23)$~K is the lowest measured value to date. However, using our optical depth measurement, the spin temperature would even be lower: $T_s=(65 \pm 17)$~K. Note that \citet{Curran2007b} reported a measurement of $T_s=60$~K in an absorbing system at $z = 0.656$ towards 3C336.

There is one galaxy candidate to the west of the background radio source, at an impact parameter of $\sim 5$~kpc. Also for this field we attempted to use the SDSS photometry to determine photometric redshifts, but the galaxy candidates are too faint to determine any useful redshift estimates.

\subsection{Low redshift \hi\ absorbers}
\begin{figure}
\begin{center}
\includegraphics[width=8.0cm,trim=1.0cm 0.5cm 0.5cm 0.5cm]{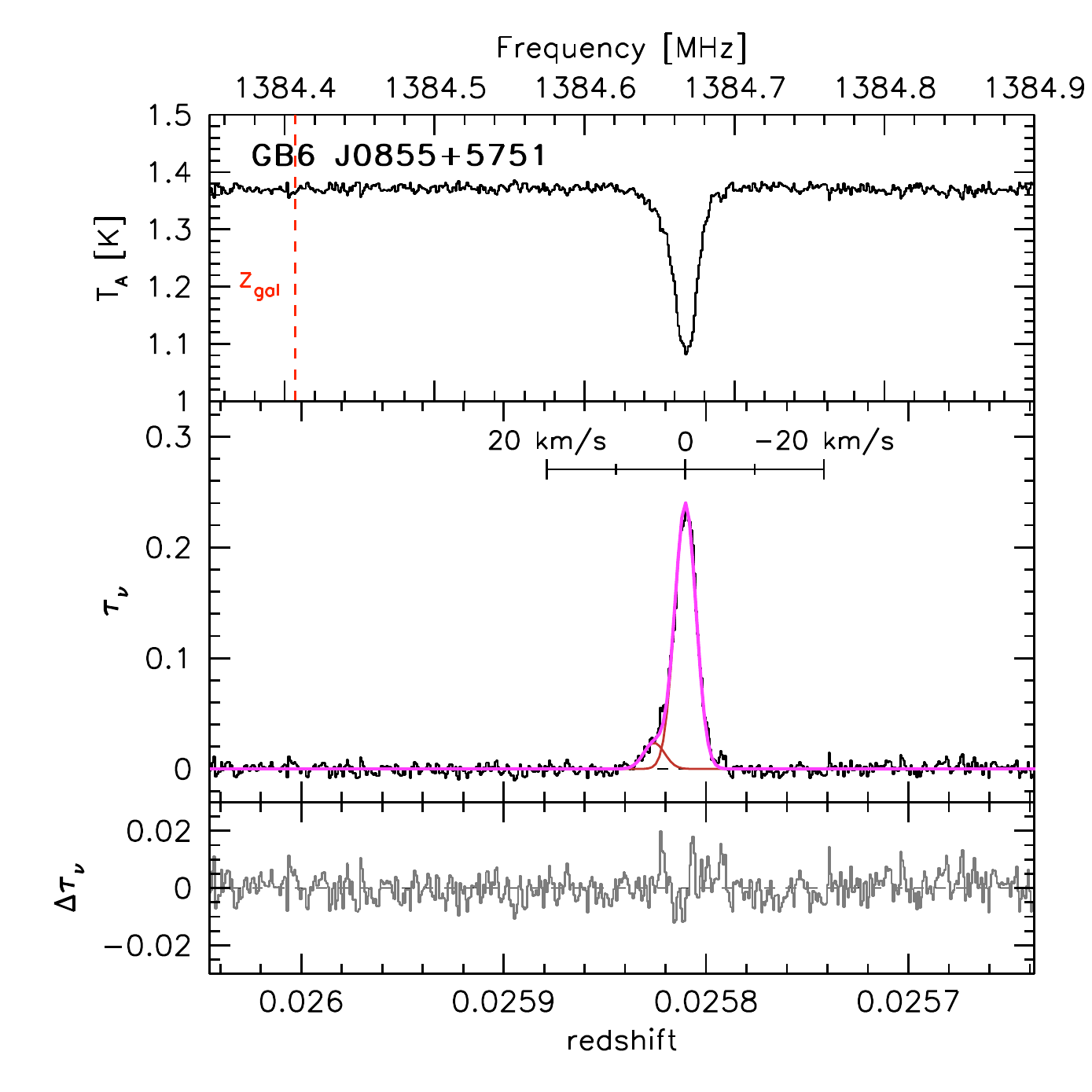}
\caption{{\em Top:\/}  GBT spectrum of the \hi\ 21-cm absorber at $z=0.026$ associated with the galaxy SDSS J085519.05+575140.7, seen against the background source GB6 J0855+5751. 
The vertical line indicates the frequency corresponding to the \hi\ line redshifted to the systemic velocity of the galaxy. {\em Middle:\/} Optical depth distribution of this spectrum. The peak optical depth is approximately 24 per cent. A double Gaussian fit to the absorption profile is shown as a solid line and the individual Gaussian components are shown by thin solid lines. {\em Bottom:\/} The residual spectrum after subtracting the fitted model. \label{0855_HI.fig}}
\end{center}
\end{figure}

\begin{figure}
\begin{center}
\includegraphics[width=8.0cm,trim=1.0cm 0.5cm 0.5cm 0.5cm]{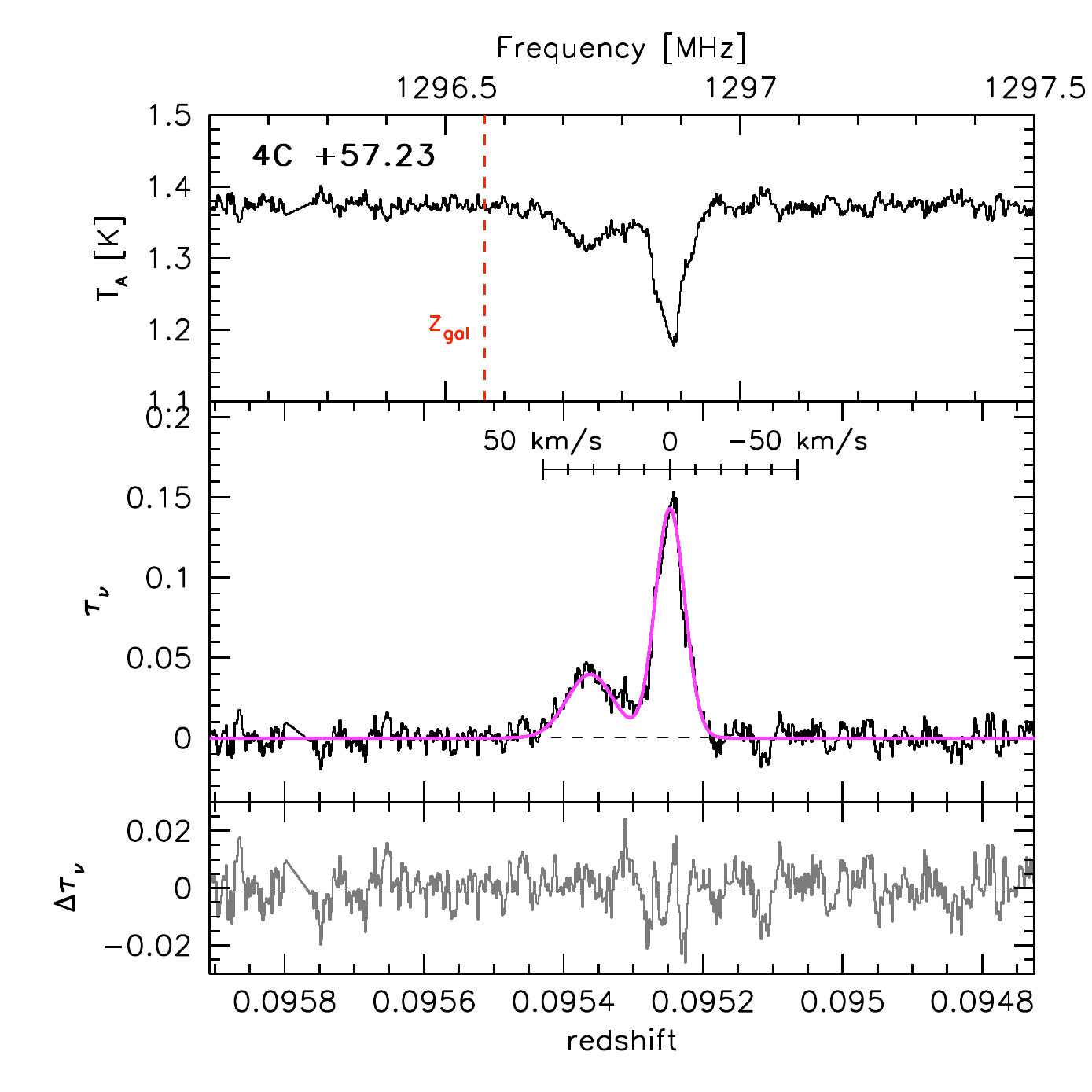}
\caption{{\em Top:\/} GBT spectrum of the \hi\ 21-cm absorber at $z=0.096$ associated with the galaxy SDSS J135400.69+565000.2, seen against the background source 4C +57.23. 
The vertical line indicates the frequency corresponding to the \hi\ line redshifted to the systemic velocity of the galaxy. {\em Middle:\/} Optical depth distribution of this spectrum. The peak optical depth is approximately 14 per cent. A double Gaussian fit to the absorption profile is shown, and the baseline is indicated by a dashed line.
{\em Bottom:\/} The residual spectrum after subtracting the fitted model. \label{4C57_HI.fig}}
\end{center}
\end{figure}

In addition to the two intermediate redshift absorbers, we identify two new strong \hi\ absorbers at very low redshifts. These are a detection at $z=0.026$ in the galaxy SDSS J085519.05+575140.7, seen against the background source GB6 J0855+5751, and a detection at $z=0.096$ in the galaxy SDSS J135400.69+565000.2, seen against 4C +57.23. The GBT \hi\ spectra of these detections are shown in Figures \ref{0855_HI.fig} and \ref{4C57_HI.fig} and multicolor SDSS images of the two fields are shown in Figure~\ref{sdssHIgal.fig}. 

\subsubsection{GB6 J0855+5751}
The absorption spectrum against GB6 J0855+5751 shows a very narrow profile, which we fit with a Gaussian with central redshift $z=0.025810\pm0.000002$, velocity dispersion $\sigma=(1.51 \pm 0.09)\,\kms$ and peak optical depth of $\tau=(0.24\pm 0.01)$. However, the profile is asymmetric at the base and a second component is required to make a satisfactory fit. We fit this second component with an offset velocity of $+4.61 \, \kms$ and a dispersion of 
$\sigma=(1.50\pm 0.20)\,\kms$. The integrated optical depth in the absorption seen against J0855 is $\tau dv=(1.02\pm 0.04)\,\kms$.

The measured width of the absorption feature places an upper limit on the amount of thermal broadening of the line. Hence, the narrowness of the absorption feature seen against J0855 can be used to derive an upper limit on the kinetic temperature of the absorbing gas. Turbulent motions could also contribute to the total measured width. The upper limit to $T_k$ can be expressed as
$$
T_k \leq \frac{m_H \Delta v^2}{k_B 8 \ln 2}  = \frac{1.2119\times 10^2 \Delta v^2}{8 \ln 2},
$$
where $k_B$ is the Boltzmann constant, $m_H$ is the mass of the hydrogen atom and $\Delta v$ is the profile width measured at half maximum. Using this equation, we find an upper limit of $T_k=275$~K for both components, which implies that we are probing the cold neutral medium (CNM). Now, equipped with a measurement of the integrated optical depth and an upper limit on the kinetic temperature, we can use Eq.~\ref{eq2.eq} to place an upper limit on the \hi\ column density of $5.0\times 10^{20}$~\icmsq, assuming the covering factor $f$ is equal to 1 and that the spin temperature is equal to the kinetic temperature. The background source GB6 J0855+5751 was tentatively identified as a Compact Symmetric Object (CSO) using 5 GHz VLBA observations by \citet{Helmboldt2007b}, which reveal two lobes separated by $\sim$55 mas. In a forthcoming paper we present new 21-cm spectral line VLBI observations of the target and discuss the small scale spatial variations of the absorption. These observations show that the absorption arises against both components and the integral optical depth varies  across the background radio source. 

\cite{Braun2012a} recently investigated the correlation between integrated optical depth and column density from emission and absorption lines studies from the literature. It was found that the relation between the two quantities, $\tau dv$ and \hi, can be fitted very well with a model with a "sandwich geometry" of cool gas with properties surrounded by layers of warm gas.  Interestingly, based on our total opacity of $\tau dv=1.02$ this model would predict a total \hi\ column density of $6.0\times 10^{20}$~\icmsq, very close to the value that we determine above.

\begin{figure}
\begin{center}
\includegraphics[width=3.5cm,trim=1.7cm 11.0cm 4.2cm 0.cm]{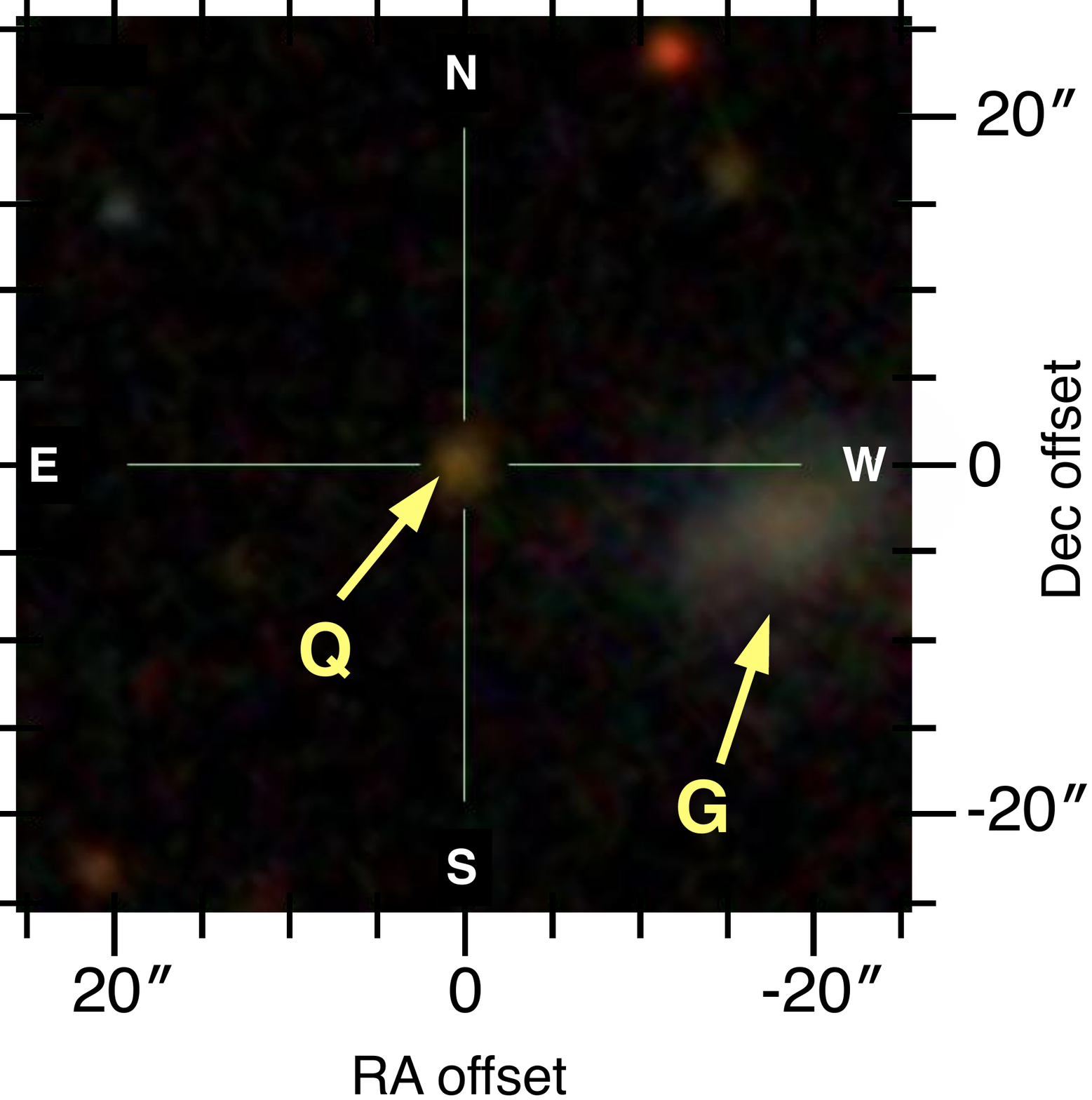}
\includegraphics[width=3.5cm,trim=0.9cm 11.0cm 5.0cm 0.cm]{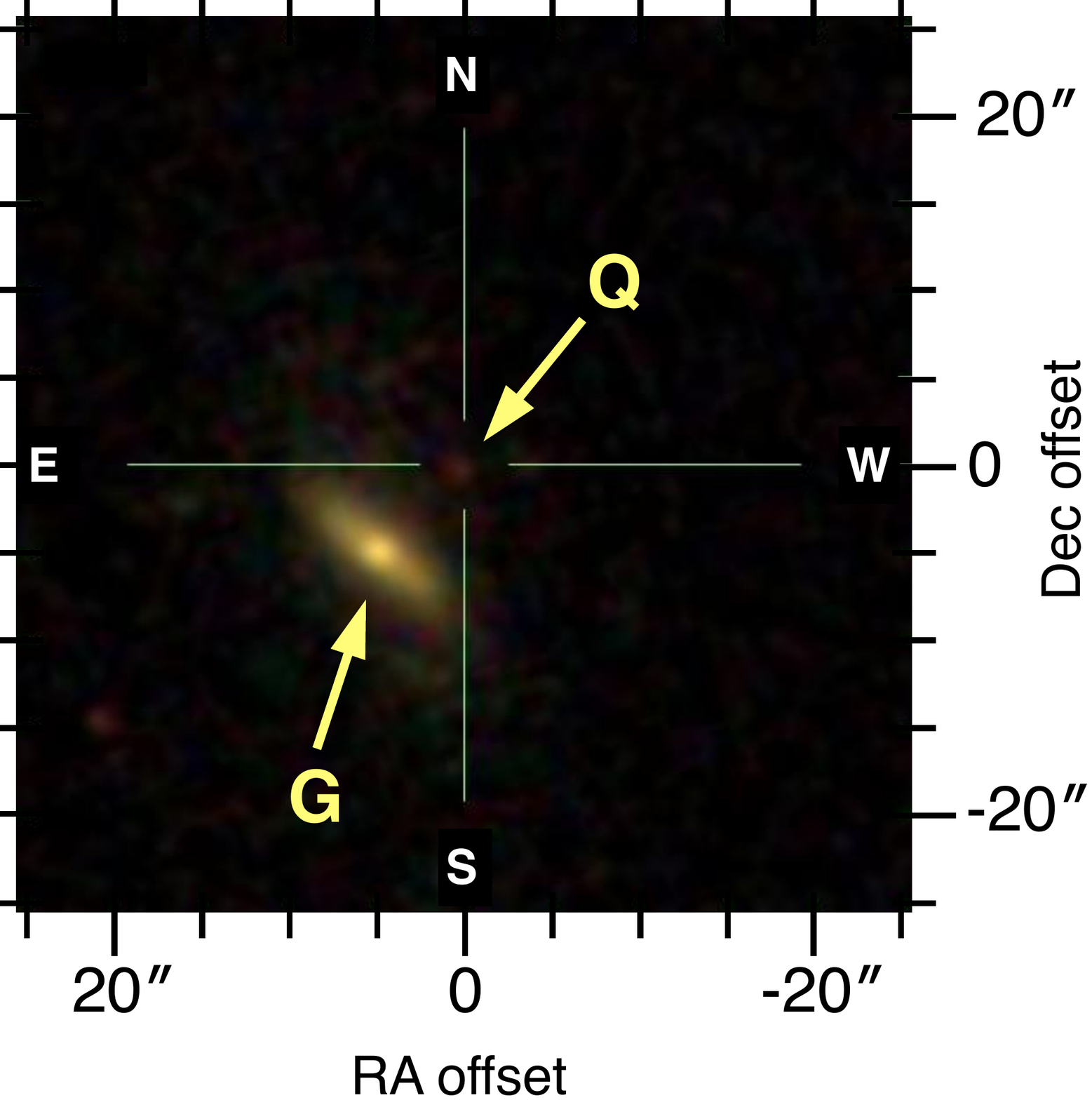}
\caption{Multicolour SDSS images of the fields around GB6 J0855+5751 (left) and 4C +57.23 (right). In both images the background source is in the centre and marked with 'Q'. The foreground galaxies can be easily identified, and are marked with 'G'. The axes are in arcsec, offset from the positions given in Table~1. The colour images are based on a combination of $g-$, $r-$, and $i-$band images, following \citet{Lupton2004a}. \label{sdssHIgal.fig}}
\end{center}
\end{figure}

The absorption profile reaches a peak at $z= 0.02581$, which corresponds to a velocity offset of $66$~\kms\ blueward of the optical redshift ($z=0.02603$) of the galaxy SDSS J085519.05+575140.7. This galaxy is a blue, low luminosity, low surface brightness system (Figure~\ref{sdssHIgal.fig}), with an absolute Petrosian $r$-band magnitude of $-17.8$~mag. The axis ratio as measured by the SDSS is $b/a=0.7 \pm 0.1$, implying an inclination of $i=45^\circ$. The angle between the major axis of the galaxy and the vector from the centre of the galaxy to the background radio source is $\phi=45^\circ$. Under the assumption that the 21-cm absorption arises in a rotating gas disk, this would imply that the rotational velocity of the disk at a radius of 9.5~kpc is equal to $66/ \sin(i)\cos(\phi)=172~\kms$. However, the expected rotational velocity of this galaxy from the Tully-Fisher relation is $V_{\rm rot}$=63 km/s \citep{Pizagno2007a}. Therefore, we conclude that the cold gas clouds we measure are probably not part of an ordered rotating gas disk, but rather are part of a warped disk or otherwise irregular gaseous structure.

From the SDSS spectrum of the galaxy, we measure $F($H$\alpha)=25.3\times 10^{-17} \rm erg s^{-1} cm^{-2}\AA^{-1}$, from which we derive a very low star formation rate of 0.0025 $\msol \rm yr^{-1}$, assuming a \citet{Chabrier2003a} initial mass function (IMF) and no correction for dust. We measure a metallicity of 12+log(O/H)=8.54, using the N2 index of \citet{Pettini2004a}. Based on the SDSS photometry, we can estimate the stellar mass of this galaxy. We apply the method detailed in \citet{Hatziminaoglou2009a}, which uses the \citet{Bertelli1994a} Simple Stellar Population (SSP) models, and find that SDSS J085519.05+575140.7 is a very low mass galaxy with a  stellar mass of $\approx 2.0 \times 10^9~\msol$ (converted to a \citet{Chabrier2003a} IMF). The stellar mass and  metallicity are consistent with the mass-metallicity relation of \citet{Tremonti2004a}. 

\subsubsection{4C +57.23}
The spectrum against 4C +57.23 has a very different shape, displaying two spectrally separated components. The stronger component has a peak optical depth of $\tau=0.14$ and can be fitted very well with a Gaussian with $\sigma=(5.8\pm {0.3})~\kms$. The weaker component reaches a peak optical depth $\tau=0.038$, and has $\sigma=(9.1\pm {0.5})~\kms$.  The integrated optical depth in the absorption seen against 4C +57.23 is $\tau dv=1.17\pm 0.08\,\kms$. After subtracting the two components, some residual structure is present, in particular in between the peaks of the two components, at $z=0.0953$. However, the signal to noise in the spectrum is not sufficiently high to fit a third component. 

The peak optical depth is seen at a velocity $70~\kms$ blueward of the optical redshift of the galaxy SDSS J135400.69+565000.2. This appears to be an edge-on spiral galaxy, probably of morphological type Sa. The SDSS spectrum of the galaxy does not show any strong emission lines. The sightline to the background radio source intercepts this galaxy along its minor axis in a region where some low surface brightness star light can be seen. 
We note that most other galaxies that have been seen in 21-cm absorption are of late morphological type, the only exception being J0849+5108 at a redshift of $z=0.3$, discovered by \citet{Gupta2013a}.

\section{Low redshift galaxy-quasar pairs}
With the two new local 21-cm absorbers reported in this paper, the total sample of intervening 21-cm absorption systems arising in low $z$ galaxies rises to 14. Table~\ref{results.tab} gives an overview of all known low redshift systems, listing the redshifts, impact parameter between the identified galaxy and the background radio source, and the integral optical depth $\tau dV$. We somewhat arbitrarily adopted a  redshift limit of $z<0.5$ to include the two systems recently presented in  \citet{Gupta2013a}, and the system at $z=0.43$ identified by \citet{Kanekar2002b}. \citet{Gupta2013a} reported a weak correlation between impact parameter and $\tau dV$, with a Spearman rank coefficient of $r=-0.3$. Adding our two new systems and two systems from the literature that \citet{Gupta2013a} did not consider (J074110.85+311154.7  and KAC2002), we find a slightly higher value of $r=-0.39$, with an associated significance level of 0.81. 

\begin{table*}
\label{results.tab}
\centering
\begin{minipage}{140mm}
\caption{Summary of all known intervening 21-cm absorption systems arising in $z<0.5$ galaxies.}
\begin{tabular}{l l l l l l l}
\hline
galaxy & redshift & impact parameter & impact parameter & $\tau dV$ & reference & \caii\ $W_0^{3934}$ \\
           &              & (arcsec) & (kpc) & (\kms) & &\\
\hline\hline
UGC~7408			& 0.0015  &  93		& 2.6 	& 1.3		& 1		& . . .\\ 
NGC~3067			& 0.0049	& 110	& 11.1 	& 0.11	& 2,3 	& $0.43 \pm 0.05$\\
ESO~1327-2041		& 0.0180	& 38.5	& 14	 	& 0.14a	& 2,4		& $0.58 \pm 0.04$  \\
J085519.05+575140.7	& 0.0258  	& 19.6 	& 9.5		& 1.02	& this work & . . .\\
NGC~6503			& 0.0270	& 18.7	& 10	 	& 0.22	& 6,2		& $0.35 \pm 0.08$\\
GQ1042+0747			& 0.0332	& 2.5		& 1.7		& 0.19 	& 5		& $<1.07$\\
0248+430				& 0.0520	& 15.0	& 15	 	& 0.26	& 7		& $1.52 \pm 0.17$\\
J163956.38+112802.1	& 0.079	& 3.6		& 5 		& 15.7	& 8 		& . . . \\
J135400.69+565000.2	& 0.0952  	& 6.8		& 11		& 3.03	& this work & . . . \\
J124157.26+633237.6	& 0.143   	& 4.4 	& 11 		& 2.90 	& 9 		& $1.01 \pm 0.11$\\
J074110.85+311154.7     	& 0.2212  & 6.0	 	& 20	 	& 0.36	& 10,11 	& . . . \\
J084958.09+510826.7	& 0.3120	& 3.0		& 13		& 0.95	& 12		& $0.60 \pm 0.18$\\
J144304+021419		& 0.3714	& $<1$	& $<5$	& 3.4		& 12		& . . .\\
KAC2002				& 0.4367 	& 2.2	 	& 11.6	& 0.75	& 13 		& . . .\\
\hline
\end{tabular}
\\{\small References: 
1: \citet{Borthakur2014a};
2: \citet{Carilli1992a};
3: \citet{Keeney2005a};
4: \citet{Keeney2011a};
5: \citet{Borthakur2010a};
6: \citet{Boisse1988a};
7: \citet{Hwang2004a};
8: \citet{Srianand2013a};
9: \citet{Gupta2010a};
10: \citet{Chengalur1999b};
11: \citet{Cohen2001a};
12: \citet{Gupta2013a}
13: \citet{Kanekar2002b}
}
\end{minipage}
\end{table*}%

Under the assumption that low redshift 21-cm absorbers arise in the gas disks of galaxies, such a correlation with a large scatter is expected. \hi\ emission line maps of local galaxies typically show a declining \hi\ column density toward larger galactocentric radii, but for a sample of \hi\ absorbers the distribution of absorption strength vs. impact parameter is a convolution of \hi\ profiles and the \hi\ size distribution function of galaxies. In \citet{Zwaan2005a} local galaxies are used to calculate the probability function of \hi\ cross-section in the column density vs. impact parameter plane for absorption selected systems. The conditional probability distribution shows that the median impact parameter $b$ at a column density of $\log \nhi=20.3$ (which defines the DLA limit) is 10~kpc, whereas the most likely impact parameter is $b=5$ kpc. 

It is interesting to compare the sample of low redshift intervening 21-cm absorption systems arising in galaxies with the likelihood distribution of \hi\ emission in the $\nhi-b$-plane. However, Equation 2 shows that for the calculation of column densities from optical depths, a measurement of the spin temperature is required, which is unknown for most of the systems in Table~\ref{results.tab}. \citet{Chengalur2013a} use simulations to show that the isothermal estimator \citep{Dickey1982a} can be used to calculate \hi\ column densities from optical depth measurements with an accuracy of a factor 2. In order to arrive at a more accurate relation between optical depth and column density, \citet{Braun2012a} used a sandwich model of cool gas, surrounded by layers of warm gas. Using their relation, and the assumption that the total velocity width of the absorption profile is 15~\kms, we can transform the $\tau dV$ values in Table~\ref{results.tab}  to column densities. This calculation uses the assumptions that the 'threshold column density', for which the optical depth approaches zero, is $N_0=1.25\times 10^{20}~\icmsq$ and the saturation column density, where the optical depth reaches infinity, is $N_\infty=7.5\times 10^{21}~\icmsq$.

\begin{figure}
\begin{center}
\includegraphics[width=10.1cm,trim=7.0cm 3.0cm 4.0cm 3.0cm]{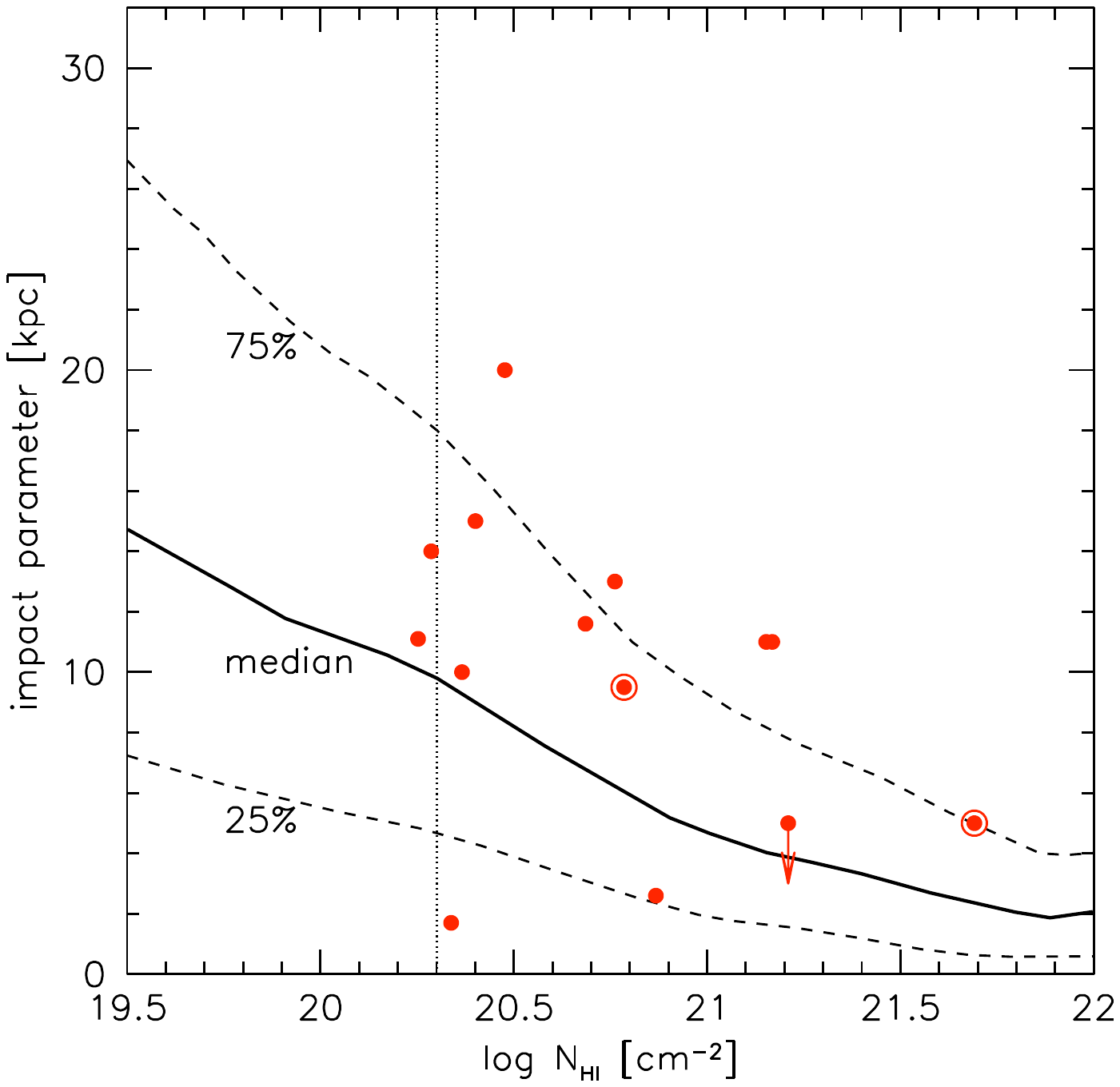}
\caption{Impact parameter vs. derived \hi\ column density (using the conversion per \citet{Braun2012a}) of low-redshift quasar-galaxy pairs. The ringed points indicate the two new low redshift absorbers reported in this paper.  The lines indicate the probability distribution functions of impact parameter derived by \citet{Zwaan2005a}, based on a large sample of local \hi\ 21-cm emission line maps. The solid line is the median impact parameter as a function of \hi\ column density, and the long-dashed lines are the 25 and 75 percentiles. The vertical line indicates the DLA column density limit. \label{rho-nhi.fig}}
\end{center}
\end{figure}

In Figure~\ref{rho-nhi.fig} we plot the measured impact parameters $b$ against these derived column densities. Also shown is the conditional probability distribution of impact parameter as a function of column density as measured by \citet{Zwaan2005a} in local galaxies. Shown are the median impact parameter and the 25 and 75 percentiles. Note that the measured distribution of absorbing systems follows the general trend that is expected from a random incidence of sight lines through local galaxies. However there is a dearth of low impact parameter systems. This effect is expected as quasar-galaxy pairs can only be identified if the angular separation between the galaxy and the background source is sufficiently large so that the two sources can be separated. Even if the background source is taken from a radio catalogue, the pair would only be identified if the separation were larger than several arcsecs, or otherwise the background source could possibly be associated with the foreground galaxy.

\citet{Reeves2015a} recently presented a survey for 21-cm absorption and emission in a sample of six nearby, gas-rich galaxies. The impact parameters between the background radio sources and the foreground galaxies ranged between 10 and 20 kpc. No \hi\ 21-cm absorption was detected in any of their galaxies, while in our sample of five galaxies, absorption was detected in two cases. Although the statistics are poor, the difference between our results and the \citet{Reeves2015a} results are probably due to the fact that {\em 1)} the continuum flux of the background sources in our study was higher, and {\em 2)} our sample includes systems with smaller impact parameters. In fact, we find that out of our sample of five, 21-cm absorption was detected in the two galaxy-quasar pairs with the smallest impact parameters ($\sim$10 kpc). \citet{Reeves2015a} also note that the detection rate of 21-cm absorption is probably strongly correlated with the structure of the background source, with compact radio sources producing higher absorption line sensitivity.

Similar relations between impact parameter and \hi\ column density of DLA galaxies have been found at higher redshifts. Using optical imaging of a large sample of DLAs, \citet{Rao2011a} find an anti-correlation between impact parameter and \hi\ column density  at redshifts $0<z<1$. They find a median impact parameter of 16.2 kpc for the DLA galaxy sample. \citet{Peroux2011a} apply the integral field unit (IFU) technique to detect redshifted H$\alpha$ emission associated with DLAs, and find indications of an anti-correlation at redshifts $1<z<2$. 
At redshifts $z>2$, \citet{Krogager2012a} combine their own observations of three DLA systems with literature results. The combined sample of 10 identifications shows a Spearman rank coefficient of $r=-0.6$ between \nhi\ and impact parameter, with an associated significance level of 93 per cent (our calculation). Using stacking, \citet{Noterdaeme2014a} find \lya\ emission in a sample of extremely strong damped \lya\ systems with 
$\nhi>0.5 \times 10^{22}~\icmsq$ at $2<z<4$. They conclude that the impact parameter of these high column density systems must be smaller than 2.5~kpc, again confirming the anti-correlation between impact parameter and \nhi\ for DLAs. Interestingly, the median impact parameter for DLA galaxies does not seem to evolve strongly. While \citet{Zwaan2005a} quote a median impact parameter of 8~kpc at $z=0$, \citet{Krogager2012a} find a value of 7~kpc at $z\sim 2.5$. Hydrodynamic simulations at $z=3$ can reproduce the anti-correlation between impact parameter and \hi\ column density  \citep{Rahmati2014a}.

\section{OH absorption line results}
In addition to our search for \hi\ absorption lines, we observed a subset of our targets in the OH 1667~MHz line. For these observations we selected the group of small impact parameter quasar-galaxy pairs and the \caii\ absorbers. The motivation to search for molecular lines in small impact parameter systems is the observation that local galaxies usually show CO emission only in their central regions.
From CO imaging of nearby galaxies \citep[e.g.,][]{Helfer2003a,Leroy2009b}, we see that the area over which CO is detectable is much smaller than the \hi\ area above the DLA limit. In addition, the regions where the CO columns are highest ($\nhtwo>$ a few $\times 10^{21}~\icmsq$) often show a depression in \hi\ column density, in many cases below the DLA limit.
Furthermore, \citet{Liszt1999a} demonstrated by comparing Galactic absorption spectra of several poly-atomic molecules with that of \hi\, that the molecules generally avoid the deepest \hi\ features and often the high column density molecules have no corresponding \hi. Presumably, in the regions where the molecular column densities are highest, most of the \hi\ has been converted to \htwo\ \citep{Schaye2001b}.

\citet{Wild2005a}  found that \caii\ absorbers have relatively high dust content and typically high \hi\ columns. In a subsequent paper \citep{Wild2006a}, they showed that the \caii\ systems have highly depleted refractory elements and high dust-to-metals ratios with values close to, or even larger than, those observed locally. These properties suggest that a substantial fraction of the \caii\ absorbers are more chemically evolved than typical DLAs. The combined results of \citet{Wild2005a, Wild2006a, Nestor2008a, Zych2009a} lead to an interpretation of the \caii\ absorbers arising in the inner parts of galaxies, preferentially regions of high gas density, with cross-sections only 20-30 per cent of those of DLAs. These arguments make them excellent candidates for molecular absorption line studies.

\citet{Liszt1999a} showed that the abundances of OH and HCO$^+$ are the best tracers of molecular hydrogen, demonstrating a near constant relation between their column densities. Arguably, OH is a more accurate indicator of \htwo\ column density than molecules observable in the millimetre regime, including the often used tracer CO.
Furthermore, in all five cases where molecular absorption has been seen in redshifted absorption line systems, corresponding OH absorption has subsequently been detected:
B3~1504+377 ($z=0.673$) and PKS~1413+135 \citep[$z=0.247,$][]{Kanekar2002a},
B~0218+357 \citep[$z=0.685,$][]{Kanekar2003a},
1830-211 \citep[$z=0.886,$][]{Chengalur1999a}, and 
PMN~J0134-0931 \citep[$z=0.765,$][]{Kanekar2005a}.

\begin{table*}
\label{OHnondetections.tab}
\centering
\begin{minipage}{100mm}
\caption{Parameters of the spectra in which OH absorption was searched for. No detections were made.}
\begin{tabular}{l l l l l}
\hline
Name & $\sigma_\tau$ & $\tau dV_{5,10} $ & $dV$ & comments\\
 & & (\kms) & (\kms) \\
\hline\hline
SBS 0846+513 	& 0.0091		& $<0.20$		&0.97 & RFI\\	
GB6 J0855+5751 	& ...			& ...			&  ... & lost in RFI\\
B3 0927+469 		& 0.0045		& $<0.11$		& 0.71 & 	RFI	\\	
4C +04.33 		& ...			& ...			& ... & lost in RFI\\
GB6 J1103+1114 	& 0.0046		& $<0.11$		& 1.51 &	\\
Q1148+387 		& ...			& ...			& ... 	& lost in RFI\\
8C 1213+590 		& 0.0032		& $<0.088$	& 1.52 &\\
B1239+606 		& 0.0039		& $<0.10$		& 1.97 &\\
PG 1241+176 		& 0.0041		& $<0.11$		& 0.68 &	lots of RFI\\
B3 1325+436 		& 0.013 		& $<0.33$		& 1.70 &\\
4C +57.23 		& 0.0025		& $<0.054$	& 0.99 &					\\
PKS 1545+21 		& 0.0019		& $<0.051$	& 1.17 &					\\
3C336 			& 0.011		& $<0.41$		& 1.66 &	strong RFI below 880.6 MHz\\
3C336-1			& 0.0049		& $<0.13$		& 1.70 &\\
PKS 2135Ð14 		& 0.00086		& $<0.025$	& 1.08 &					\\	
PKS 2330+005 	& 0.0041		& $<0.081$	& 1.05 &					\\
\hline
\end{tabular}
\end{minipage}
\end{table*}%

Table~\ref{OHnondetections.tab} lists the systems that were observed in the OH line. The observational set-up was the same as that described in section 3, and data reduction was identical to what was done for the HI observations. 
Unfortunately, none of the systems we observed were detected in absorption in the 1667~MHz OH  line. We should note that seven of the spectra were severely affected by RFI. In particular for the low-redshift quasar-galaxy pairs that were observed in the frequency range of 1500 to 1600~MHz, no useful data could be obtained. 

 Table~\ref{OHnondetections.tab} summarises the upper limits to the integral optical depth $\tau dV$ that can be derived from our measurements. To convert the optical depth limits into OH column densities, we use the equation given by \citep{Liszt1996a}:
\begin{equation}
N_{\rm OH}=2.24 \times 10^{14} \times T_{\rm ex} \int \tau_{1667} dV \, [{\rm cm}^{-2}],
\end{equation}
where $T_{\rm ex}$ is the excitation temperature in Kelvin.
The value of $T_{\rm ex}$ is somewhat uncertain. As argued by \citet{Kanekar2003a}, the excitation temperature in redshifted absorbers is typically higher than $T_{\rm CMB}(1+z)$. In the Milky Way, a typical value of $T_{\rm ex}=10~\rm K$ is found in dark clouds. Lacking better constraints, we also assume here that  $T_{\rm ex}=10~\rm K$. Finally, to convert OH column densities into \htwo\ column densities we adopt the relation from \citet{Liszt1999a}: $\nhtwo=1.0 \times 10^7 \times N_{\rm OH}$.

\begin{figure}
\begin{center}
\includegraphics[width=10.1cm,trim=7.0cm 3.0cm 4.0cm 3.0cm]{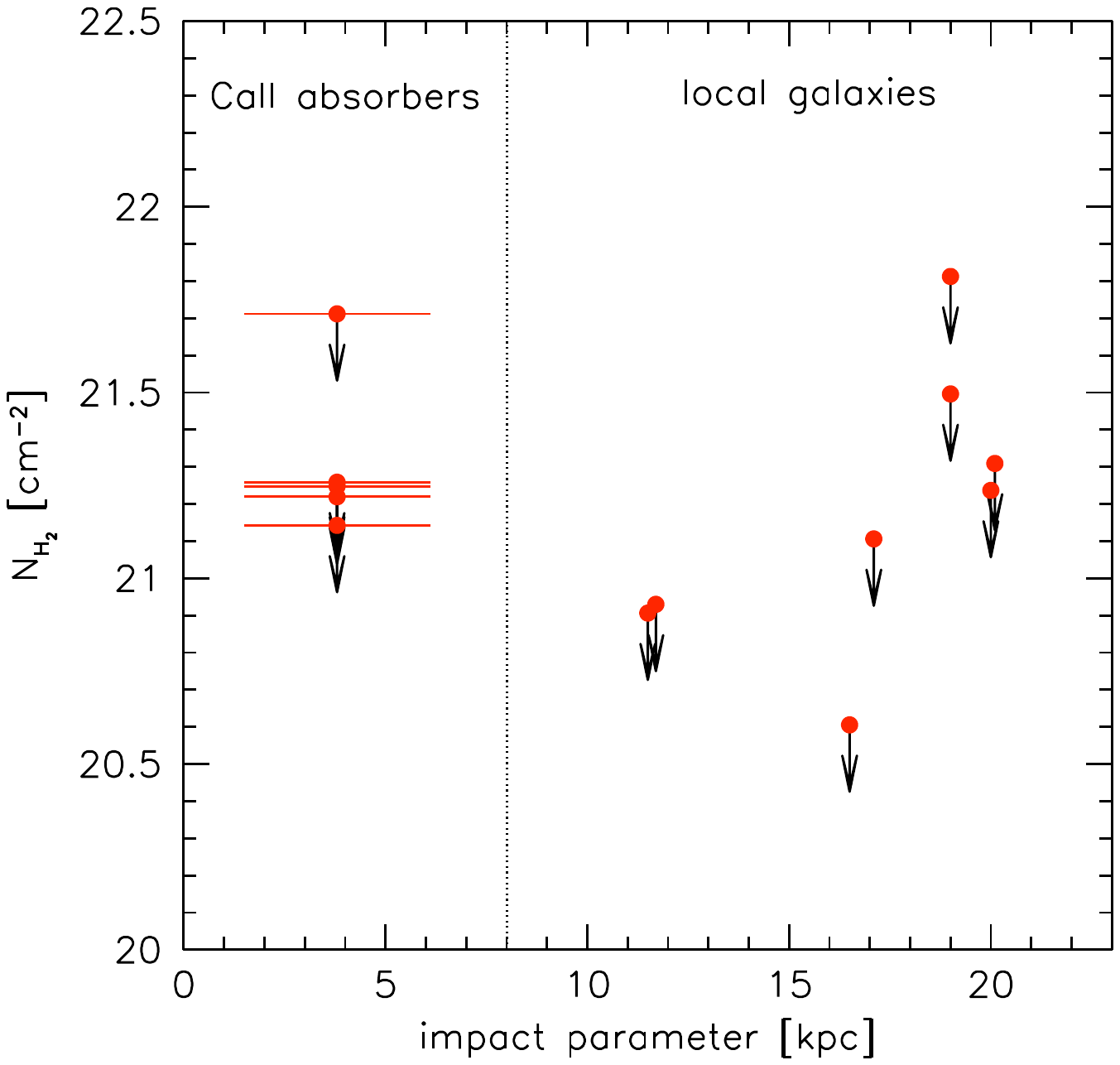}
\caption{Upper limits on the \htwo\ column densities, derived from the OH absorption non-detections. For the local galaxy-quasar pairs, the upper limits are plotted against their impact parameter. The \caii\ absorbers are plotted at a fixed impact parameter of 4 kpc. \label{OHupperlimits.fig}}
\end{center}
\end{figure}

In Figure~\ref{OHupperlimits.fig} we plot the upper limits to the \htwo\ column densities as a function of impact parameter. We also show the measurements in the five \caii\ systems, for which we have no measurement of the impact parameter because the host galaxies of the \caii\ absorption have not been identified. Our $\nhtwo$ upper limits are typically a few $\times 10^{21} \icmsq$. It is interesting to compare these values to the cumulative  distribution function of impact parameters of \htwo\ cross-section-selected systems, as derived by \citet{Zwaan2006a}. There it is shown that, based on the analysis of CO maps of local galaxies, it is expected that \htwo\ column densities of $\log \nhtwo>21$ are only detected if the impact parameter is less than 10~kpc. Our non-detections are consistent with these expectations.

Also for the \caii\ systems at intermediate redshifts, we reach upper limits of $\nhtwo \approx 2 \times 10^{21} \icmsq$. At $z=0$ we would expect to detect these column densities at a median impact parameter of 3~kpc \citep{Zwaan2006a}. As stated earlier, 
\citet{Wild2005a} interpret the \caii\ absorbers as the inner parts of galaxies, with cross-sections only 20-30 per cent of those of DLAs. Adopting a median impact parameter of low-redshift DLAs of 7.6~kpc \citep{Zwaan2005a}, we would expect that the median impact parameter of \caii\ systems is therefore $\sim 3.8$ kpc. Using this line of reasoning, we could expect to have detected OH absorption in one or two \caii\ systems in our small sample of five. Now that larger samples of \caii\ absorbers are available, it would be very interesting to repeat the experiment of detecting molecular gas -- OH or other molecules such as CO -- in these systems.

\section{Conclusions}
We present the combined results of several Green Bank Telescope surveys for intervening \hi\ and OH absorption at low and intermediate redshifts ($0<z<1.2$). We obtained good quality data for 17 out of 24 systems -- observations of the remaining seven systems were severely affected by radio frequency interference. The total sample of observed systems consists of \caii\ absorbers, \mgii\ absorbers,  apparent quasar-galaxy pairs with small impact parameters at low redshift, and \mgii\ and \lya\ absorbers with identified host galaxies. Our conclusions are the following:

\begin{enumerate}
\item Out of five low redshift ($z<0.17$) quasar-galaxy pairs, we detected strong \hi\ 21-cm absorption in two systems. The maximum 21-cm optical depths are 14 and 24 per cent, respectively. Both absorption profiles are very narrow, the narrowest has a velocity dispersion of only 1.5~\kms, which puts an upper limit on the kinetic temperature of $T_k < 270$~K. Combining these two systems with 21-cm absorbers from the literature, we measure a tentative anti-correlation between impact parameter and integral optical depth. Using a sandwich model for the cool gas, we convert optical depth measurements into column densities. The resulting correlation between impact parameter and \hi\ column density is in good agreement with what is expected if 21-cm absorbers arise in the gas disks of galaxies that are seen in 21-cm emission.
\item Of the sample of five \caii\ and  \mgii\ systems for which good data were obtained, we measure 21-cm absorption in one \mgii\ system, and one \caii\ system, both at $z\approx 0.6$. Since some of the \caii\ absorbers are also \mgii\ absorbers, in total three \mgii\ systems have been observed. The detection rate is in agreement with previous studies of \mgii\ systems. Given the indications that \caii\ systems are on average higher \hi\ column density absorbers, we would expect a higher detection rate for these systems, but unfortunately, the statistics are too poor to draw any conclusions on this. 
\item No OH absorption was detected in any of our systems. Comparing our results with the expected strong anti-correlation between column density and impact parameter based on what is observed in local galaxies, our non-detections are fully consistent with the expectations. To increase the sample of known intervening molecular absorption lines, even smaller impact parameter systems ($b<10$~kpc) need to surveyed. When larger samples of \caii\ absorbers become available, we recommend observations of molecular absorption lines in these systems with radio or millimetre bright background sources. 
\end{enumerate}

By following up known optical quasar absorption line systems in the radio, the total sample of 21-cm absorbers has increased steadily over the last few decades. However, the next big step in the study of 21-cm absorption systems is expected with the appearance of wide field blind surveys with upcoming instruments such as MeerKAT \citep{Booth2009a}, ASKAP \citep{Deboer2009a}, and APERTIF \citep{Oosterloo2009a}. For example, the ASKAP FLASH\footnote{First Large Absorption Survey in \hi;\\ http://www.physics.usyd.edu.au/sifa/Main/FLASH/} survey will search for 21-cm absorption features in approximately 150,000 sight-lines to background radio sources brighter than 50 mJy at 800 MHz. The full southern sky survey will increase the number of known intervening 21-cm absorbers in the range $\nhi\sim10^{19}$~\icmsq\ to $10^{22}$~\icmsq\ between $0.4 <z<1.0$ with a factor of 20. The very first pilot results of FLASH \citep{Allison2015a} resulted in the detection of a new absorber at $z=0.44$ and demonstrate the ability to efficiently perform wide-field \hi\ absorption surveys. Surveys with APERTIF and MeerKAT will probe lower down the column density distribution function, but will cover smaller areas. Together, the result of these surveys will lead to a much better understanding of the global evolution of cold gas in the $z<1$ Universe, but will also turn up many interesting instances of galaxies at intermediate redshifts in which the ISM can be studied in great detail. The examples presented in this paper illustrate the type of intervening absorption systems that will be detected routinely in these surveys, and eventually with the Square Kilometre Array\footnote{www.skatelescope.org}. 

\section*{Acknowledgments}
We thank the anonymous referee for their useful comments on the manuscript.
We also thank Stephane Arnouts for help with running the Le Phare photometric redshifts code and
Evanthia Hatziminaoglou for measuring the stellar mass of the galaxy SDSS J085519.05+575140.7 from 
SDSS photometry.
The data presented in this paper have been collected as part of programs
GBT06A-061,
GBT06A-063,
GBT06B-022,
GBT06C-053, and
GBT06C-055.
We are grateful to the NRAO staff for their support with the GBT observations.

\small
\bibliographystyle{mn2e-new}
%\bibliography{mn-jour,all}
\bibliography{/Users/mzwaan/REFERENCES/zwaanreferences}

\null\vspace{4cm}
\label{lastpage}

\end{document}